%% file: phain-tf-text.tex
\documentclass[lettersize,journal]{IEEEtran}

\usepackage{amsmath,amsfonts}
\usepackage[algoruled,linesnumbered]{algorithm2e}
    \SetKw{Break}{break}
\usepackage{array}
\usepackage{textcomp}
\usepackage{stfloats}
\usepackage{url}
\usepackage{verbatim}
\usepackage{graphicx}
\usepackage{multirow}
\usepackage{footmisc}
\usepackage{cite}
\usepackage{upgreek}
\usepackage{tikz}
\usetikzlibrary{external}
\usetikzlibrary{positioning}
\usetikzlibrary{arrows.meta}
\usepackage{pgfplots}
\pgfplotsset{compat=1.18}

\usepackage{adjustbox}

\usepackage{hyperref}

\input{defs}
\SetKw{Or}{or}
\begin{document}
\title{Audio Inpainting in Time-Frequency Domain\\ with Phase-Aware Prior}
\author{Peter Balušík, Pavel Rajmic, and Ondřej Mokrý
\thanks{The work was supported by the Czech Science Foundation (GA\v{C}R) Project No.\,23-07294S. The authors
thank V.\,Kovanda for plotting routines used in Figure~\ref{fig:demo.phase.correction}, and K.\,Yatabe for valuable comments on an earlier draft of the manuscript.
Special thanks also go to all participants of the listening test, 
and to the anonymous reviewers for insightful comments and suggestions.

The authors are with the SPLab at Department of Telecommunications,
FEEC, Brno University of Technology, 616 00 Brno, Czech Republic (email:
\href{mailto:230531@vut.cz}{230531@vut.cz};
\href{mailto:rajmic@vut.cz}{rajmic@vut.cz};
\href{mailto:Ondrej.Mokry@vut.cz}{Ondrej.Mokry@vut.cz}).
}}

\markboth{IEEE TRANSACTIONS ON SIGNAL PROCESSING, VOL. XX, XXXX}%
{Balušík \MakeLowercase{\textit{et al.}}: Audio inpainting in the time-frequency domain with phase-aware prior}


\maketitle

\begin{abstract}
We address the problem of time-frequency audio inpainting, where the goal is to fill missing spectrogram portions with consistent information. 
Despite recent advances, existing approaches still face limitations in both reconstruction quality and computational efficiency.
To bridge this gap, we propose a~method that utilizes a phase-aware signal prior which exploits estimates of the instantaneous frequency. 
An optimization problem is formulated and solved using the generalized \mbox{Chambolle--Pock} algorithm.
The proposed method is evaluated against other \mbox{time-frequency} inpainting methods, specifically a~deep-prior audio inpainting neural network, the autoregression-based \mbox{approach} known as Janssen-TF, and a~sparsity-driven baseline.
For short gap durations,
all regularization-based methods achieve comparable results in terms of SNR, which is due to the redundancy of the particular time-frequency transform used.
On longer gaps, the proposed approach performs comparably to Janssen-TF.
Nonetheless, in terms of the perceptual quality (both objective and subjective), the proposed method consistently outperforms existing methods. 
In addition, the reconstructions are obtained with a~substantially reduced computational cost compared to alternative methods.
\end{abstract}

\section{Introduction}
\label{sec:intro}
\IEEEPARstart{A}{udio} inpainting is the task of replacing missing (or corrupted) parts of a digital audio
recording \cite{Adler2012:Audio.inpainting}.
Such parts are usually referred to as \textit{unreliable} \cite{MokryRajmic2020:Inpainting.revisited,TanakaYatabeOikawa2024:PHAIN}.
In the case of audio inpainting in the time domain, they take the form of randomly missing samples or the form of \emph{gaps}, which is the case when unreliable samples are concentrated in one or several places. 
Unreliable samples can originate from transmission errors, faulty equipment during recording, or data corruption.

Several methods exist for the inpainting of such gaps: those based on the sparsity of the time-frequency representation of audio \cite{MokryZaviskaRajmicVesely2019:SPAIN,Mokry2022:Audio.inpainting.NMF,TanakaYatabeOikawa2024:PHAIN,ToumiEmiya2018:Sparse.Non.Local.Inpainting,Lieb2018:Audio.Inpainting,MokryRajmic2020:Approximal.operator,TaubockRajbamshiBalasz2021:SPAINMOD},
autoregression-based methods such as \cite{javevr86,Etter1996:Interpolation_AR},
similarity-based methods \cite{Bahat_2015:Self.content.based.audio.inpaint,Perraudin2018:Similarity.Graphs},
and recent methods employing deep learning \cite{Greshler2021:Catch.A.Waveform,Lemercier2024:Diffusion.Audio.Restoration}.
Sparsity-based and autoregression-based methods are usually successful at inpainting of shorter gaps (tens of milliseconds), while deep learning-based methods are often able to fill in larger gaps (hundreds of milliseconds up to several seconds).

\begin{figure} [htb]
    \centering
    \includegraphics[width=0.44\textwidth]{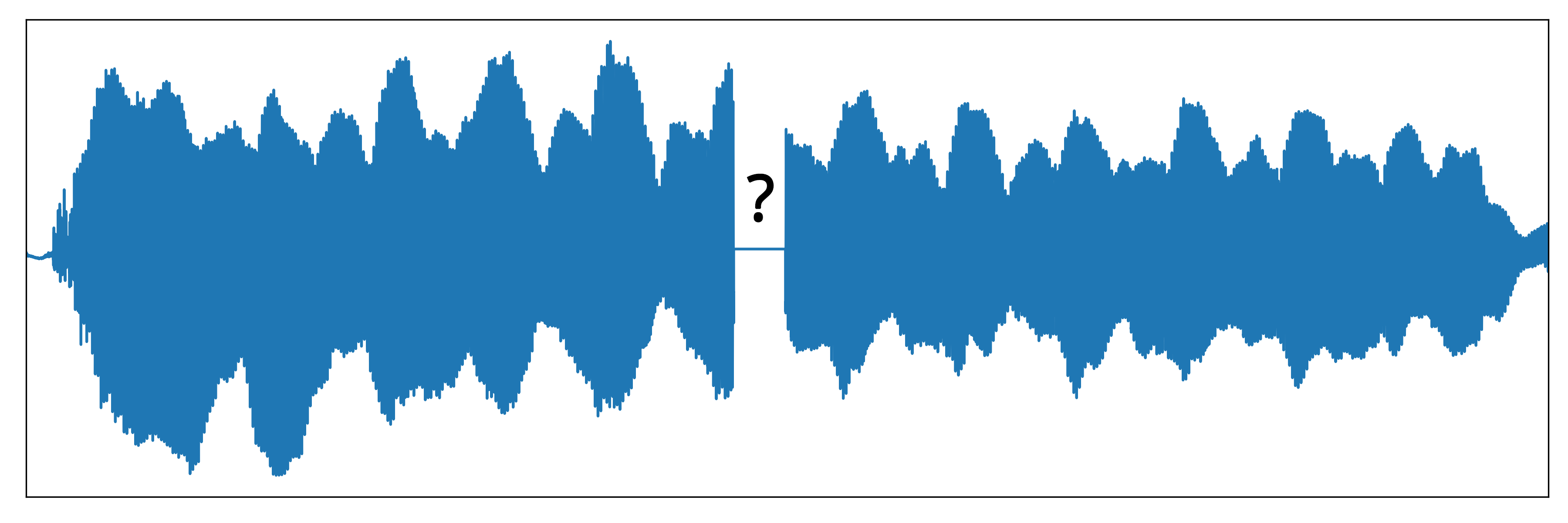}\\ \vspace{5pt}
    \includegraphics[width=0.34\textwidth]{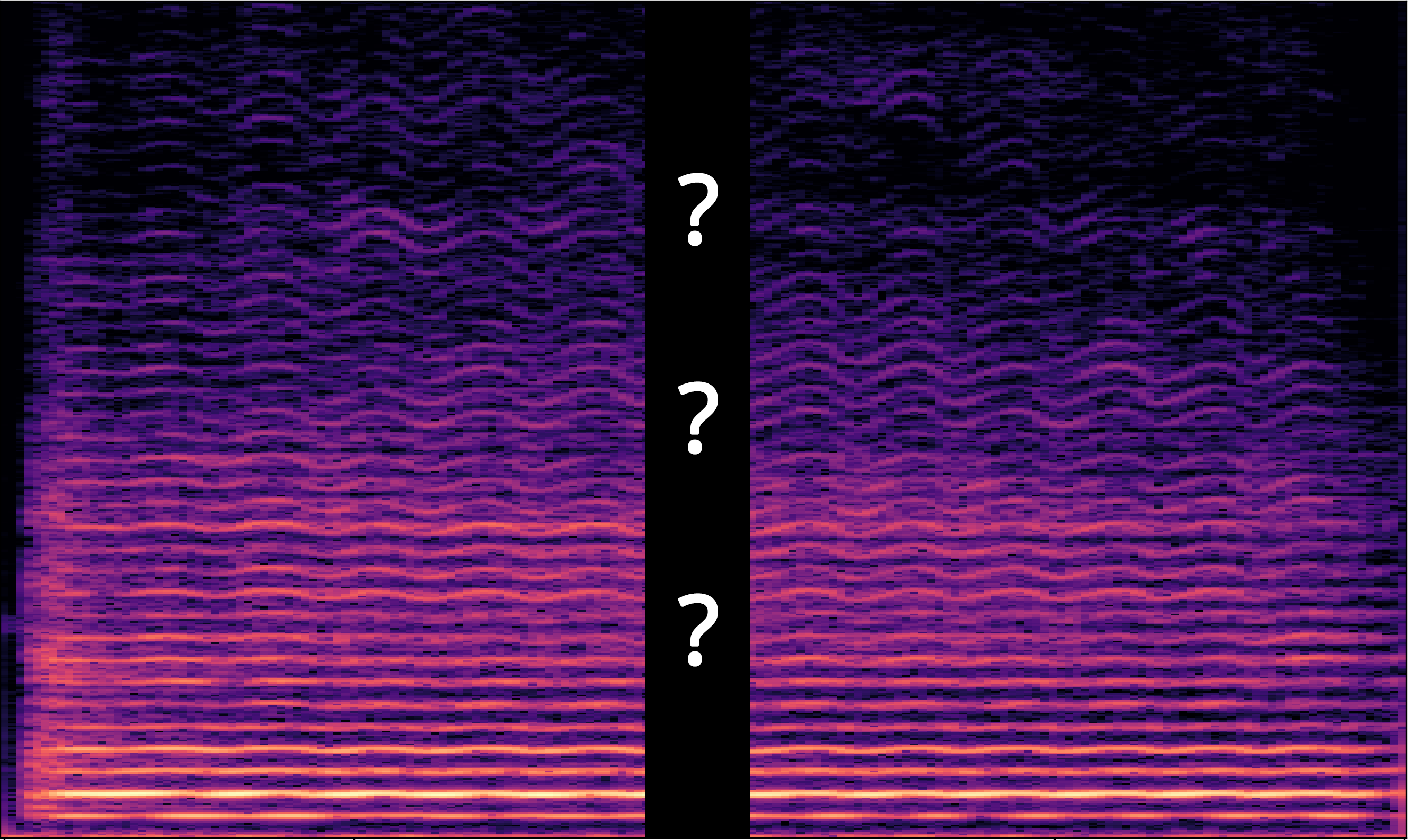}
    \caption{Audio inpainting problem in the time domain (top) and in the time-frequency domain (bottom).}
    \label{fig:gaps}
\end{figure}
The above-mentioned methods fall into the category of audio inpainting in the time domain.
Recently, the idea of audio inpainting in the time-frequency (TF) domain, or \textit{spectrogram inpainting}, arose.
This is motivated by the use of short-time signal spectra in many audio codecs, such as MP3, AAC, CELT \cite{Bosi2003:DigitalAudioCoding},
and by the suitability of TF representations in deep-learning-based applications.
Indeed, several such methods have been proposed for spectrogram inpainting \cite{Marafioti2019:Context.encoder,
Marafioti2021:GACELA,Miotello2023:Deep.Prior.Inpainting.Harmonic.CNN, Liu2023:MAID}.
Lately, a~method based on autoregression called Janssen-TF \cite{MokryBalusikRajmic2025:Spectrogram.inpainting} has been shown to be the superior method both objectively and subjectively.
In the case of spectrogram inpainting, a~\emph{gap} refers to missing TF coefficients rather than missing samples;
see Fig.~\ref{fig:gaps} for an illustration.
The length of the gap is expressed as the number of missing columns of the spectrogram.
Importantly, gaps
(and related inpainting problems)
in the two domains are \textit{not equivalent};
see \cite{MokryBalusikRajmic2025:Spectrogram.inpainting,RajmicBartlovaPrusaHolighaus2015:Acceleration.support.restriction}
and a~brief explanation in Sec.~\ref{sec:T.and.TF.inpainting.nonequivalence}.

The spectrogram in Fig.~\ref{fig:gaps} presents a signal representation that is substantially sparser than the signal in the time domain.
By sparsity, we refer to the property of a vector or matrix in which only a relatively small number of its entries attain significant values, while the majority of elements are zero or close to zero.
For natural signals, suitable transformations can usually be found leading to sparse coefficients.
In audio, the short-time Fourier transform (STFT) is often utilized;
sparsity-based audio inpainting methods then stem from the observation that the time-frequency coefficients are more or less sparse.
Consequently, the simplest audio inpainting method minimizes the plain sparsity of the spectrogram elements, where the $\ell_1$-norm is often used as a convex approximation, making the problem computationally feasible.
More sophisticated methods promote particular structures (mostly horizontal) in the TF representation via a~suitable prior.

While inpainting methods based on $\ell_1$-norm minimization have become successful, they come with a number of disadvantages.
The main one is the so-called energy loss problem~\cite{MokryRajmic2020:Inpainting.revisited},
i.e., the amplitude of the reconstructed audio signal decreases towards the center of the gap.
This is given by the structure of the STFT and by the inherent properties of the $\ell_1$-norm.
Another problem is the lack of attention to the temporal connections 
between sinusoidal components in the TF domain~\cite{TanakaYatabeOikawa2024:PHAIN}. 
Both of these problems were recently addressed by the instantaneous Phase-Corrected Total Variation (iPCTV) prior, originally introduced in \cite{YatabeOikawa2018:ipctv}, and then applied to the time-domain audio inpainting in
\cite{TanakaYatabeOikawa2024:PHAIN}. 
The authors of~\cite{TanakaYatabeOikawa2024:PHAIN} proposed
PHase-aware Audio INpainter (PHAIN) to solve audio inpainting problems in the time domain.
One type of their method, known as U-PHAIN, has proven to be the superior method, surpassing even the state-of-the-art Janssen algorithm~\cite{javevr86,MokryRajmic2025:Inpainting.AR}.

This paper integrates the idea of the instantaneous frequency in the problem of inpainting in the time-frequency domain.%
\footnote{The concept has been presented as an invited poster at the SampTA 2025 conference; see the abstract at \url{https://openreview.net/forum?id=j0DHbpebCU}.
No actual paper has been published, however.}
Therefore, our developed method is abbreviated as
\mbox{U-PHAIN-TF}.
The proposed method is compared with the Janssen-TF method
\cite{MokryBalusikRajmic2025:Spectrogram.inpainting}
and the Deep Prior Audio Inpainting (DPAI) method
\cite{Miotello2023:Deep.Prior.Inpainting.Harmonic.CNN}, using objective evaluation and a~MUSHRA-style listening test.
In addition, U-PHAIN-TF is compared with a sparsity-driven baseline to isolate the impact of the iPCTV prior.
The DPAI method is based on the deep prior approach \cite{UlyanovVedaldiLempitsky2020:Deep.Image.Prior-journal}, 
adapted to inpainting audio instead of images.
Even though the paper presented satisfactory results, the method is highly computationally demanding. 
The \mbox{Janssen-TF} method has been shown recently to outperform DPAI in terms of both objective and subjective evaluation, while requiring a~lower computational effort \cite{MokryBalusikRajmic2025:Spectrogram.inpainting}.
We challenge the reconstruction efficiency and computational demand of Janssen-TF with our newly proposed method, \mbox{U-PHAIN-TF}.     

The paper is organized as follows.
In Section \ref{sec:preliminaries}, the STFT, the problem of spectrogram inpainting, and the generalized Chambolle--Pock algorithm (GCPA) \cite{Condat2023:Proximal.splitting.algorithms} are discussed.
Next, in Section \ref{sec:phain}, the concept of phase-aware prior is summarized. 
The new method, U-PHAIN-TF, is proposed in Section~\ref{sec:uphaintf}. 
Section \ref{sec:refMethods} briefly characterizes the reference methods used in the experiments,
which are presented in Section \ref{sec:expe}.
A~thorough discussion of the results is followed by the conclusion in Section \ref{sec:conclude}.

\section{Preliminaries} \label{sec:preliminaries}
\subsection{Short-time Fourier Transform (STFT)} \label{sec:stft}
The discrete-time short-time Fourier transform, also known as the discrete Gabor transform or just STFT, is widely used in time-frequency (TF) signal processing \cite{Grochenig2001:Foundations.T-F.analysis}. 
For a
signal $\x\in\Rset^L$ of length $L\in\Nset$ and a window function $\g\in\Rset^L$\!
of effective length $w\ll L$,
the STFT is defined as%
\begin{equation} \label{eq:STFTdefnition}
    X[m,n]=(\ana_\g \x)[m,n]= \sum_{l=0}^{L-1}x[l]\cdot g[l-an]  \cdot \eul^{-\jmag 2\uppi ml/M}\!,
\end{equation}%
where \(l\) indexes elements of $\x$,
while \(m \in \{0,1,\dots,M-1\}\)
and \(n \in \{0,1,\dots,N-1\}\) denote the frequency and time indices, respectively. 
The hop size $a \in \Nset$ is chosen such that $aN=L$, and the index $l-an$ is evaluated  modulo~$L$.
In this article, the number of frequency channels is set equal to the effective window length ($M = w$). 
Note that the definition~\eqref{eq:STFTdefnition} corresponds to an STFT with a~frequency-invariant phase \cite{Prusa2015:phase.Convention,Yatabe2019:representation.complex.spec},
and solely this STFT type will be used throughout this paper.
The STFT of the signal $\x$ is indicated by $\X=\ana_\g \x$, where $\X\in\Cset^{M\times N}$ is a complex \textit{spectrogram}.%
\footnote{Note that with $\x$ being real-valued, the spectrogram is symmetric with respect to the 
DC component.
Thus, in practice, the frequency dimension can be halved.
Note also that, in much of the literature, the term \qm{spectrogram} is used to denote only the magnitude part, in contrast to our current work.}
The elementwise absolute value and the argument can be used to separate the spectrogram into its \textit{magnitude} and \textit{phase} parts.

Whereas formula \eqref{eq:STFTdefnition} corresponds to the analysis STFT operator
$\ana_\g\colon\Rset^L\rightarrow\Cset^{M\times N}$\!,
its adjoint
$\syn_\g\colon\Cset^{M\times N}\rightarrow\Rset^L$\!,
often called the inverse STFT or a~synthesis STFT, is defined as
\begin{equation} \label{eq:STFTdefnition.synthesis}
     x[l] =(\syn_\g \X)[l]= \sum_{m=0}^{M-1}\sum_{n=0}^{N-1}X[m,n]\cdot g[l-an] \cdot \eul^{\jmag2 \uppi ml/M}\!.
\end{equation}

In this paper, exclusively STFT operators satisfying the conditions on the so-called Parseval tight frames will be utilized. They satisfy the following properties \cite{christensen2008,MokryRajmic2020:Inpainting.revisited}
$$
\syn_\g\ana_\g = \Id,\quad
\norm{\ana_\g}^2= 1, 
$$
where $\Id$ denotes the identity operator (in this context, the identity on $\Rset^L$).

Note that the STFT is typically used in a~redundant regime, i.e., \mbox{$MN > L$}.
In such a case, the null space of $\syn_\g$ is a~vector space of a~nonzero dimension.
As a consequence, 
a~single spectrogram can be computed from one signal only,
whereas an identical signal can be synthesized from multiple matrices of size $M\times N$
(only one of which lies in the range space of $\ana_\g$, however).
%
In this article, the term spectrogram will be used for any complex matrix of size $M\times N$, for instance for a~matrix created from $\ana_\g \x$ by turning some elements to zero.
To distinguish the cases, a~spectrogram that can be obtained by the STFT analysis will be called consistent.

\subsection{Sparsity-based Inpainting}
\label{sec:Sparsity.based.inpainting}
As mentioned in Section \ref{sec:intro}, sparsity-based inpainting methods promote particular structures in the TF representation via a~suitable prior
$\phi$, often corresponding to an optimization problem of the form
\begin{equation}
    \label{eq:analysisModel}
    \argmin_{\x\in\Rset^L}\ \phi(\ana_\g\x)\quad\text{s.t.}\quad \x\in\sigset,
\end{equation}
where $\sigset$ is the set of all feasible time-domain signals 
\begin{equation} \label{eq:setOfALLSigs}
        \sigset = \{\x\in\Rset^{L}\ |\ \mask\odot\x=\mask\odot\xorig\}.
\end{equation}
Here, $\mask$ is a~binary mask for time signals, indicating unreliable signal samples with zero and reliable elements with one, and $\odot$ is the Hadamard (elementwise) product.
As an example, the $\ell_1$-norm of a~vectorized matrix can be used in place of $\phi$ to obtain a~simple time-domain inpainting formulation \cite{MokryRajmic2020:Inpainting.revisited},
but more sophisticated priors can be involved
\cite{TanakaYatabeOikawa2024:PHAIN,MokryRajmic2020:Approximal.operator}.


\subsection{Spectrogram Inpainting}
\label{sec:sgram.inpainting}
Having the STFT and the spectrogram defined, the actual time-frequency inpainting problem can now be formalized.
Similar to Section \ref{sec:Sparsity.based.inpainting}, for audio inpainting in the time-frequency domain, the following applies:
\begin{itemize}
    \item
        $\Xorig\in\Cset^{M\times N}$ is the ground truth (original) spectrogram obtained as $\Xorig =\ana_\g\xorig$, where $\xorig$ is the ground truth signal.
    Neither the ground truth signal nor the spectrogram are known in practice. In this paper, they are used solely as tools to compute the audio metrics in the objective evaluation, see Section \ref{sec:objectiveEval}.  
    \item
        $\Tfmask\in\{0,1\}^{M\times N}$ is a binary mask analogous to $\mask$, but for spectrograms. 
    The indices of the unreliable spectrogram elements are assumed known.
    As in \cite{Marafioti2021:GACELA,Miotello2023:Deep.Prior.Inpainting.Harmonic.CNN,MokryBalusikRajmic2025:Spectrogram.inpainting}, the focus is on the case when whole columns of the spectrogram are missing.
    In such a case, $\Tfmask$ consists exclusively of zero columns or columns of ones.
    \item
        $\Xcor\in\Cset^{M\times N}$ is the observed (corrupted) spectrogram modeled using the binary mask $\Tfmask$ such that it holds $\Xcor=\Tfmask\odot\Xorig$\!.
    \item
        Consequently $ \coefset$ is defined as the set of all feasible spectrograms 
    \begin{equation} \label{eq:setOfALLSpectrograms}
        \coefset = \{\X\in\Cset^{M\times N}\ |\ \Tfmask\odot\X=\Tfmask\odot\Xorig\},
    \end{equation}
    i.e., a~feasible spectrogram must match the observation.
\end{itemize}
With this notation, we can write a generic TF-inpainting reconstruction problem as follows:
\begin{equation}
    \label{eq:analysisModelSpecgram}
    \argmin_{\x\in\Rset^L}\ \phi(\ana_\g\x)\quad\text{s.t.}\quad \ana_\g\x\in\coefset.
\end{equation}

Note that although the case of entire missing columns of the spectrogram is considered in this paper, generalization to other cases is straightforward.
Indeed, $\coefset$ may represent an arbitrary subset of spectrogram coefficients, where the selection is governed by the mask $\Tfmask$ in \eqref{eq:setOfALLSpectrograms}.
This also encompasses cases in which entire frequency bins are missing, for example.
From a computational perspective,
all such cases are handled by the same projection operator, see Section~\ref{sec:uphaintf}.

\subsection{Time-domain Inpainting and Time-frequency-domain Inpainting Are Not Equivalent}
\label{sec:T.and.TF.inpainting.nonequivalence}
From Fig.~\ref{fig:gaps},
the temptation might arise to conclude that the problem can be solved interchangeably in either domain.
Although the formulations in Sections~\ref{sec:Sparsity.based.inpainting} and \ref{sec:sgram.inpainting}
appear strikingly similar, the two problems are not the same in the sense that a~problem posed in one domain cannot be solved in an equivalent manner in the other domain.
The underlying reason lies in the structure of the time-frequency transform:
a~window $\g$ is employed in both the analysis STFT \eqref{eq:STFTdefnition}
and the synthesis STFT \eqref{eq:STFTdefnition.synthesis}.
Consequently, a gap (i.e., a zero-valued segment) in the time domain does not map to a gap in the time-frequency domain (i.e., zero columns),
and, conversely, a gap in the time-frequency domain does not correspond to a~gap in the time domain.
The relationship is depicted in Fig.~\ref{fig:Nonequivalent.gaps}.

For more detailed illustrations of the STFT geometry, see~\cite{RajmicBartlovaPrusaHolighaus2015:Acceleration.support.restriction}.

\begin{figure}
    \centering
    \scalebox{.95}{%
    \input{tex/test_grahs}%
    }
    \vspace{-2mm}%
    \caption{%
    Illustration of how gaps relate between the time and time-frequency domains.
    Top: The STFT image of a time-domain gap of length 2048 does not appear as a~clean, empty region in the TF domain, but rather as a blurred one.
    In this example, we use an STFT with
    parameters $a=512$ and $w=2048$.
    Consequently, while seven spectrogram columns are affected, only a~single column is exactly zero, because the successive time-shifts of the window $\g$ in~\eqref{eq:STFTdefnition} overlap by 75\,\%.
    Bottom: A similar effect occurs in the reverse direction: after synthesis, the 4-column long TF-domain gap leads to a~gradually fading leakage in the time domain.
    Note that the bottom spectrogram is not consistent, whereas the top spectrogram is consistent, since it lies in the range of $\ana_\g$.
    }
    \label{fig:Nonequivalent.gaps}
\end{figure}

\subsection{Generalized Chambolle--Pock Algorithm} \label{sec:gcpa}
Proximal algorithms are efficient tools for solving convex minimization problems.
The Chambolle--Pock algorithm (CPA), also known as the primal-dual algorithm \cite{ChambollePock2011:First-Order.Primal-Dual.Algorithm,Condat2023:Proximal.splitting.algorithms},
can be used to solve optimization problems involving one linear operator, such as problem \eqref{eq:analysisModel},
see also \cite{MokryRajmic2020:Inpainting.revisited}.
The \emph{generalized} Chambolle--Pock algorithm (GCPA) \cite{Condat2023:Proximal.splitting.algorithms} is an extension of the CPA in two directions:
it solves problems involving a~linear operator in each of the terms,
precisely, problems of the form
\begin{equation} \label{eq:gcpaOptimization}
    \min_{\x}f(K\x) + g(L\x)+\inner{\x}{\mathbf{c}},
\end{equation}
and the algorithm itself utilizes extrapolation to accelerate convergence. 
A weak convergence to the solution of the above problem is guaranteed by choosing step sizes
$\tau,\sigma,\eta>0$
such that \(\tau\cdot\sigma\cdot\norm{L}^2\leq1\) and \(\tau\cdot\eta\cdot\norm{K}^2\leq1\); see \cite[Theorem~6.2]{Condat2023:Proximal.splitting.algorithms}.

\section{Phase-Aware Prior} \label{sec:phain}
As previously noted, the sparsity of the time-frequency representation can be promoted via $\ell_1$-based penalties, which in turn gives rise to the energy loss issue \cite{MokryRajmic2020:Inpainting.revisited}.  
Bayram~\cite{bayram2013:simple_prior} proposed a simple prior that forces the phase evolution along spectrogram rows to behave predictably.
Although his prior effectively addresses the energy loss problem, it does not provide high-quality solutions.
As an improvement, the instantaneous Phase-Corrected Total Variation (iPCTV)
\cite{YatabeOikawa2018:ipctv} was designed.
This penalty function is a~fundamental building block of phase-aware audio inpainting methods, such as PHAIN \cite{TanakaYatabeOikawa2024:PHAIN}, as well as of our proposed method.
The iPCTV is defined in the TF domain and it stems from the assumption that an audio signal consists of multiple but distinguishable short-time sinusoids, and from the observation that a~pure sinusoid has a~predictable phase in the corresponding spectrogram.
The iPCTV uses the concept of the instantaneous frequency (IF) as an augmentation of  \cite{bayram2013:simple_prior}, allowing adaptation of the penalty to the actual signal content.
Components of the iPCTV are discussed in the following paragraphs.

\subsection{Phase Correction}
\label{sec:phasecorrection}
\begin{figure}
    \centering
    \includegraphics[width=.96\linewidth,viewport=0 210 367 386,clip]
    {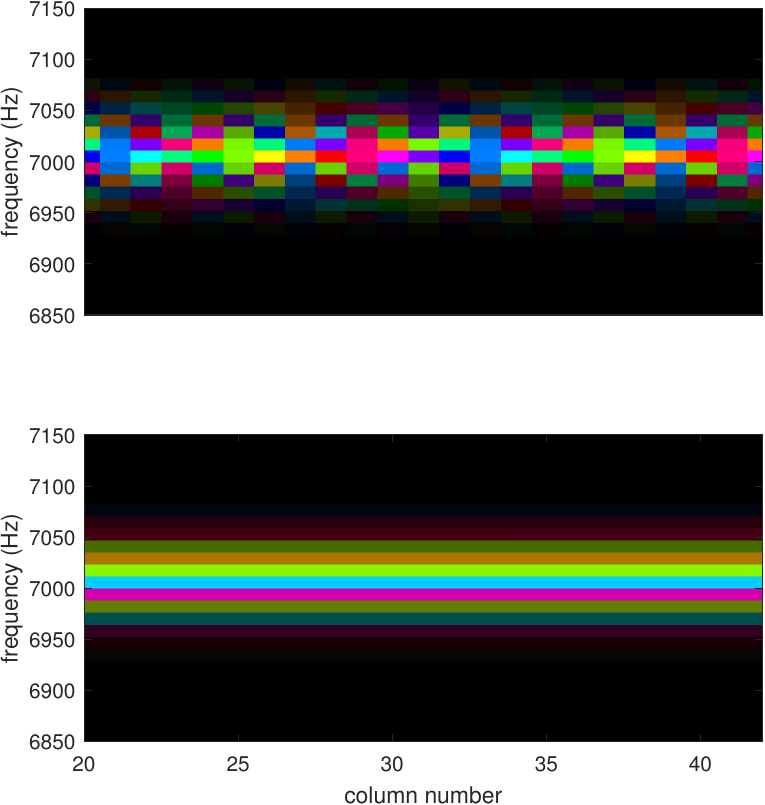}\\[-1ex]%
    \includegraphics[width=.96\linewidth,viewport=0 0 367 170,clip]
    {pdf/fig_demo_phase_correction.pdf}
    \caption{%
    The top subplot presents a tightly zoomed-in view of the spectrogram $\ana_{\g}\x$, where $\x$ consists of a single
    sinusoid at 7\,kHz.
    The bottom subplot displays the same area after phase correction, i.e., $\R_{\insFreq{\x}}\ana_{\g}\x$.
    In this visualization, the hue represents the phase of each spectrogram coefficient, while lightness reflects its magnitude.
    Hence, it is clear that $\R_{\insFreq{\x}}$ fully compensates the phase rotation and renders the evolution along the time axis constant.}
    \label{fig:demo.phase.correction}
\end{figure}

Let \(x[l]= \eul^{\jmag2\uppi(\mu+\delta)l/M}\) be a complex exponential where \(\delta\in\mathbb{R}\) stands for a frequency shift from the center frequency of the frequency bin \(\mu\in\{0,1,\dots, M-1\}\).
As shown in \cite{YatabeOikawa2018:ipctv,TanakaYatabeOikawa2024:PHAIN},
the following neighborhood relation holds for the STFT that was defined in \eqref{eq:STFTdefnition}:
\begin{equation} \label{eq:phaseRotationG}
    (\ana_\g \x)[m,n+1 ] = \eul^{\jmag2\uppi(\mu+\delta -m) a/M}\cdot (\ana_\g \x)[m,n].
\end{equation}
The phase values in the $\mu$-th frequency bin can be corrected (compensated) by a~rotation factor inverse to $\eul^{\jmag2\uppi(\mu+\delta -m) a/M}$\!, if $\delta$ is known \cite{TanakaYatabeOikawa2024:PHAIN}. 
The relative instantaneous frequency $\delta$ of
a~signal~$\x$ is estimated elementwise through
\begin{equation} \label{eq:instantPHAIN}
     \omega_{\x}[m,n] =- \Im \left[\dfrac{(\ana_{\g'}\x)[m, n]}{(\ana_\g \x)[m, n]}\right],
\end{equation}
where $\Im(\cdot)$ denotes the imaginary part and $\g'$ is the time-direction derivative of the window $\g$, see \cite{augfla95:reassign,TanakaYatabeOikawa2024:PHAIN}.
Together, these considerations lead to the following phase correction operator on the spectrogram $\X\in\Cset^{M \times N}$%
\cite{TanakaYatabeOikawa2024:PHAIN}:
\begin{equation} \label{eq:phaseCorrection}
    (\R_{\insFreq{x}}\X)[m,n] = \eul^{-\jmag2\uppi a\sum_{t=0}^{n-1}\omega_\x[m,t]/M}X[m,n].
\end{equation}
Composition of such an elementwise operator with the analysis STFT yields a~linear transform called instantaneous phase-corrected STFT, which is defined for a~signal $\x$ as 
$\R_{\insFreq{\x}}\ana_{\g}\x$~\cite{TanakaYatabeOikawa2024:PHAIN}.
A simple illustration of how the described phase correction operates on a~harmonic signal is shown in Fig.~\ref{fig:demo.phase.correction}.
Note that the IF employed in this transform does not need to be calculated directly from $\x$, which can be exploited depending on the application. 
Note that a formula slightly different from \eqref{eq:instantPHAIN}
is utilized in practice to avoid numerical instability when STFT coefficients are (close to) zero
\cite{YatabeOikawa2018:ipctv,Yatabe2019:representation.complex.spec},
as seen at line~\ref{alg:line:IFup} of Algorithm~\ref{alg:gcpaUPHAINTF}.

\subsection{Total Variation}
In an ideal case of a~single-sinusoidal signal $\x$, the TF coefficients of $\R_{\insFreq{\x}}\ana_{\g}\x$ are constant over time, see Fig.~\ref{fig:demo.phase.correction}.
In other words, the temporal variation of the phase-corrected STFT is zero across all frequency bins.
This observation motivates the introduction of
the time-directional variation operator
$\D\colon\Cset^{M \times N}\rightarrow \Cset^{M\times (N-1)}$,  defined as \cite{TanakaYatabeOikawa2024:PHAIN}
\begin{equation}
    (\D \X)[m,n]=X[m,n]-X[m,n+1],
\end{equation}
where $\X\in\Cset^{M \times N}$ is a~spectrogram.
In the ideal case, $\D$ would yield a zero matrix once applied to the phase-corrected spectrogram.
Therefore, the natural last part of the iPCTV is the  $\ell_1$-norm of the resulting matrix, as in \cite{bayram2013:simple_prior}.

\subsection{Penalty Function: iPCTV}
The iPCTV penalty $\iPCTVwhole$ is a~combination of the just described components.
Through the use of the temporal total variation on the phase-corrected spectrogram, this functional attains a~low value for signals composed of sinusoids with slowly changing amplitudes.
Thus, when used as a~penalty, sinusoidal components are retained, while components whose phase evolution cannot be approximated by linear functions (random noise as an extreme case) are rejected \cite{TanakaYatabeOikawa2024:PHAIN}.

\subsection{U-PHAIN for Time-domain Inpainting}
Based on the above considerations, the focus of PHAIN is solving the optimization problem~\cite{TanakaYatabeOikawa2024:PHAIN} 
\begin{equation} \label{eq:PHAINoptimizationT}
    \argmin_{\x\in\Rset^L}\lambda\iPCTVwhole+\ind{\sigset}(\x),
\end{equation}
i.e., seeking a feasible time-domain signal $\x$
that yields minimal possible iPCTV penalty.
Here, $\iota_{\sigset}(\x)$ is the indicator function of the convex set $\sigset$,
allowing us to write the inpainting problem in an unconstrained form,
defined as
\begin{equation} \label{eq:indicatorSpectrograms}
\iota_{\sigset}(\x) =
\begin{cases}
     \makebox[10mm][l]{$0$} \text{if} \ \x \in \sigset\\
     \makebox[10mm][l]{$\infty$} \text{if} \ \x\not\in \sigset\,.
\end{cases}
\end{equation}
Problem \eqref{eq:PHAINoptimizationT} clearly relies on the discussed instantaneous frequency estimate $\insF$.
Fixing $\insF$ to the value obtained by \eqref{eq:instantPHAIN} from the corrupted signal $\xcor$\!, and numerically solving this convex problem via the CPA 
corresponds to a~basic algorithm abbreviated B-PHAIN by the authors of~\cite{TanakaYatabeOikawa2024:PHAIN}.

Since $\xcor$\! contains gaps, this approach typically results in suboptimal phase correction and, consequently, a suboptimal reconstruction.
This issue is handled by a variant of PHAIN, abbreviated \mbox{U-PHAIN}, which incorporates a regular
IF update~\cite{TanakaYatabeOikawa2024:PHAIN}.
Accordingly, the U-PHAIN algorithm consists of an outer and an inner loop.
The inner loop solves the above optimization problem
(i.e., B-PHAIN with a few iterations of the CPA).
The outer loop updates the IF $\insF$ based on the current intermediate B-PHAIN solution, and this procedure is iterated until satisfactory convergence is reached.
Note that due to this type of update, the final U-PHAIN solution may no longer correspond to a~convex optimization formulation.

\subsection{Related Work Utilizing the Instantaneous Frequency}
Formula \eqref{eq:instantPHAIN} for the estimation of the IF is well established and has been widely employed in the signal processing literature. However, to the best of our knowledge, prior to the introduction of the iPCTV in \cite{YatabeOikawa2018:ipctv}, it had primarily been utilized to enhance the readability of time-frequency signal representations, such as the spectrogram \cite{augfla95:reassign} and synchrosqueezed transforms \cite{AugerFlandrin2013:Reassignment.Synchrosqueezing}.

\section{Phase-Aware Spectrogram Inpainting}
\label{sec:uphaintf}

Having defined the iPCTV penalty function, we now proceed to formulate the time-frequency domain audio inpainting problem.
Similar to the time-domain case \eqref{eq:PHAINoptimizationT}, the functional to optimize consists of the penalty and the feasibility constraint:
\begin{equation} \label{eq:optimizationGCPAphain}
\argmin_{\x\in\Rset^L}\lambda\iPCTVwhole+\ind{\coefset}(\ana_{\g}\x).
\end{equation}
Here, $\coefset$ is a~convex set of all feasible spectrograms defined in \eqref{eq:setOfALLSpectrograms},
and $\ind{\coefset}$ is the corresponding indicator function.
As before, the formulation aims to recover a~time-domain signal; however, in this case, feasibility constraints are imposed in the time-frequency domain.

Similarly to \eqref{eq:PHAINoptimizationT},
the optimization problem \eqref{eq:optimizationGCPAphain} includes a~parameter $\lambda>0$.
Traditionally, this scalar would be referred to as a regularization or balancing parameter of the problem.
Nevertheless, from a theoretical perspective, $\lambda$ has no effect on the minimizer of \eqref{eq:optimizationGCPAphain}.
As will be demonstrated in Sec.\,\ref{sec:hyperparameters.choice}, its choice can still strongly impact the resulting solution,
with the convergence rate being one of the aspects influenced.
Hence, $\lambda$ can be viewed as an algorithmic parameter rather than a standard regularization parameter.

Once 
the IF estimate $\insF$ is fixed,
problem \eqref{eq:optimizationGCPAphain} is convex and thanks to its structure, it can be solved using the
GCPA introduced in Section~\ref{sec:gcpa}.
When the following assignments are made in \eqref{eq:gcpaOptimization},
\begin{equation*}
    f=\ind{\coefset},\ g =\lambda\normOne{\cdot},\ K=\ana_\g,\ L=\D\R_\insF\ana_\g,\ \mathbf{c}=\mathbf{0},
\end{equation*}
it results in the inner loop of Algorithm~\ref{alg:gcpaUPHAINTF}.
In the algorithm,
the adjoint operators to $K$ and $L$ are used, which are easy to obtain.
The algorithm also requires knowledge of proximal operators of the conjugate functions $f^*$ and~$g^*$\!, i.e., operators $\prox{\eta f^*}$ and $\prox{\sigma g^*}$, see \cite{Condat2023:Proximal.splitting.algorithms}.
An operator
$\prox{ f^*}$ can be computed at the same cost as $\prox{f}$ thanks to the Moreau identity \cite{MokryRajmic2020:Inpainting.revisited, Condat2014:Generic.proximal.algorithm}
\begin{equation} 
    \prox{\eta f^*}(\x) = \x-\eta\cdot\prox{(f/ \eta)}(\x/{\eta}),
\end{equation}
for $\eta>0$. 
In our setting, we invoke the
identity to obtain 
\begin{equation}
    \prox{\eta \ind{\coefset}^*}(\X)=\X - \eta\cdot\proj{\coefset}(\X/\eta),
\end{equation}
where $\X$ is a spectrogram and $\proj{\coefset}$ represents the projection onto the convex set $\coefset$:
\begin{equation} \label{eq:projectionSpectrograms}
    \proj{\coefset}(\X) =\Tfmask\odot\Xcor +(1-\Tfmask)\odot\X,
\end{equation}
which effectively replaces the reliable TF coefficients in the input spectrogram with those from $\Xcor$\!, while leaving the unreliable coefficients unaltered
\cite{MokryRajmic2020:Inpainting.revisited}.
The proximal operator of $g^*$ is given by
\begin{align} 
        \prox{\sigma (\lambda\normOne{\cdot})^*}(\X) 
    &=\X -\sigma\cdot\soft{(\lambda/\sigma)}(\X/\sigma) \label{eq:soft1}\\
    &=\X - \soft{\lambda}(\X), \label{eq:soft2}
\end{align}
where $\soft{\lambda}$ is the soft-thresholding operator, i.e., the proximal operator of the $\ell_1$-norm \cite{Chartrand2014:shrinkage.mappings,MokryRajmic2020:Inpainting.revisited}, given by
\begin{equation} \label{eq:softThresholding}
     \soft{\lambda}(\X) = \sgn(\X)\odot\max(|\X|-\lambda,0),
\end{equation}
where $\sgn(\cdot)$ is the sign function.
The proximal operator of $g^*$ is the place where the value of $\lambda$ becomes relevant.
\begin{algorithm}[tb]
\caption{U-PHAIN-TF: using GCPA to solve the minimization problem \eqref{eq:optimizationGCPAphain}.}
\label{alg:gcpaUPHAINTF}
\small
Input a peak-normalized corrupted spectrogram $\Xcor_\text{norm}$. \label{alg:line:norm}

Choose \(\tau,\sigma,\eta > 0\) such that 
$\tau\cdot\sigma\cdot\norm{\D\R_{\insF}\ana_{\g}}^2\leq1$
and $\tau\cdot\eta\cdot\norm{\ana_\g}^2\leq1$; \(\alpha=1\); $\beta=10^{-10}$\!.

Initialize variables $\x^{(1)}\!=\syn_\g\Xcor_\text{norm}$,
$\Y^{(1)}\!=\mathbf{0}$, and $\Z^{(1)}\!=\mathbf{0}$.


Set output variable $\hat{\x}^{(1)}\!=\x^{(1)}$\!; set $I$, $J$, and threshold $\varepsilon>0$. 
\\[1mm]
    \For{$j=1,\dots,J$}{
     $\insF^{(j)}\!=\!-\Im[\ana_{\g'}\hat{\x}^{(j)}\overline{\ana_{\g}\hat{\x}^{(j)}}\!\!\oslash\!( |\ana_\g \hat{\x}^{(j)}\!|^2\!+\! \beta\max |\ana_\g \hat{\x}^{(j)}\!|^2)]$\label{alg:line:IFup}\\
    \For{$i=1,\dots,I$}{
        \(\sR = \!\Y^{(i)}\!+\eta\cdot\ana_\g\bigl(\x^{(i)}\!-\tau\cdot(\syn_{\g}\Rad_{\insF^{(j)}} \!\Dad\Z^{(i)}\!+\ana_\g^*\Y^{(i)})\bigl)\)\label{alg:line:Yupdate}\\
        \(\Y^{(i+\frac{1}{2})}=\sR-\eta\cdot\proj{\coefset}(\sR/\eta)\)\\
        \(\x^{(i+\frac{1}{2})} =\x^{(i)} - \tau\cdot( \syn_{\g}\Rad_{\insF^{(j)}} \Dad\Z^{(i)}+\syn_{\g}\Y^{(i+\frac{1}{2})})\)
        \\
         \(\sQ=\Z^{(i)}+ \sigma\cdot\D\R_{\insF^{(j)}}\ana_{\g}(2\cdot\x^{(i+\frac{1}{2})}-\x^{(i)})\) \label{alg:line:Qupdate}\\
       \(\Z^{(i+\frac{1}{2})}=\sQ - \soft{\lambda}(\sQ)\)
       \\
         \(\x^{(i+1)} = \x^{(i)} + \alpha\cdot(\x^{(i+\frac{1}{2})}-\x^{(i)})\)\\
         \(\Y^{(i+1)} = \Y^{(i)} + \alpha\cdot(\Y^{(i+\frac{1}{2})}-\Y^{(i)})\) \\
         \(\Z^{(i+1)} = \Z^{(i)} + \alpha\cdot(\Z^{(i+\frac{1}{2})}-\Z^{(i)})\)\\
        }
    \uIf{warmRestart}
    {
        $\x^{(1)}\!=\x^{(I+1)}$, $\Y^{(1)}\!=\Y^{(I+1)}$, $\Z^{(1)}\!=\Z^{(I+1)}$ \label{alg:line:warmInit}
    }
    \Else{
        $\x^{(1)}\!=\syn_\g\Xcor_\text{norm}$, $\Y^{(1)}\!=\mathbf{0}$, $\Z^{(1)}\!=\mathbf{0}$ \label{alg:line:coldInit}
    }
    \(\hat{\x}^{(j+1)} = \x^{(I+1)}\) \\
    \If{$\norm{\hat{\x}^{(j+1)}-\hat{\x}^{(j)}}_2<\varepsilon$}
    {
        \Break
    }
    }
    \Return{$\hat{\x}$ \Or $\proj{\coefset}(\ana_\g \hat{\x})$} \label{alg:line:return}
\end{algorithm}

By additionally incorporating an update step for the IF estimate, we obtain the final proposed Algorithm~\ref{alg:gcpaUPHAINTF} called instantaneous-frequency-Update PHase-aware Audio INpainter in the Time-Frequency domain (U-PHAIN-TF).
Its inner loop ($i=1,\ldots,I$) uses the GCPA to solve \eqref{eq:optimizationGCPAphain}.
The outer loop ($j=1,\ldots,J$) provides a~regular IF update (line \ref{alg:line:IFup} of Alg.~\ref{alg:gcpaUPHAINTF}). 
The symbol $\oslash$ at line \ref{alg:line:IFup} denotes the elementwise division.
Note that the input spectrograms are peak-normalized (line~\ref{alg:line:norm} of Alg.~\ref{alg:gcpaUPHAINTF}), as detailed later in Sec.~\ref{sec:implement}.

Despite fulfilling the sufficient conditions ensuring the inner loop convergence,
it should be noted that this is guaranteed only for $\insF$ remaining fixed in \eqref{eq:optimizationGCPAphain}.
Since $\insF$ regularly updates after finishing the inner loop, no general convergence guarantee can be provided for the algorithm as a whole.
Nevertheless, no issues of that kind were observed in our experiments.

One primal and two dual variables are used in the GCPA;
in our case, $\x\in\Rset^L$\!, \(\Y\in \Cset^{M\times N}\)\! and~\(\Z\in\Cset^{M \times (N-1)}\).
The initialization of these variables, and their update once the inner iteration loop has completed, is controlled by the \texttt{warmRestart} parameter.
If enabled, the variables are warm-started from the solution obtained in the previous inner loop (line \ref{alg:line:warmInit}). 
Otherwise, a~cold-start initialization is performed, resetting the dual variables to zero (line \ref{alg:line:coldInit}).
Additionally, during the algorithm, two temporary variables $\sR\in\Cset^{M \times N}$\! and $\sQ\in\Cset^{M\times (N-1)}$\!
are utilized along with a~relaxation parameter $\alpha\in(0,2)$. 
In particular, we consider the setting $\alpha=1$, as in \cite{TanakaYatabeOikawa2024:PHAIN}.
Since problem \eqref{eq:optimizationGCPAphain} is posed in the time domain,
 the algorithm naturally returns the estimate $\hat{\x}$.
Nevertheless, since the task is to complete missing spectrogram columns,
one may instead desire to obtain the reconstructed spectrogram $\hat{\X}=\proj{\coefset}(\ana_\g \hat{\x})$, 
as seen at line~\ref{alg:line:return}.

Note that phase-aware spectrogram inpainting can be formulated in several different ways beyond \eqref{eq:optimizationGCPAphain}.
We have also analyzed a problem formulated solely in the time-frequency domain; see Appendix \ref{app:alternative}.
However, such an approach turned out to be less efficient than Alg.~\ref{alg:gcpaUPHAINTF}.
The results indicate that it is important when the variables of the GCPA alternate between the time domain and the TF domain.
We also examined alternative thresholding operators to soft thresholding and alternative norms to the $\ell_1$-norm; see Appendix \ref{app:thresholding}.
Nevertheless, soft thresholding performed the best in terms of objective metrics, suggesting that the use of the $\ell_1$-norm
on phase-corrected spectrograms
is beneficial. 

\section{Reference Methods} \label{sec:refMethods}
Several approaches to spectrogram inpainting have recently emerged, including Deep Prior Audio Inpainting \cite{Miotello2023:Deep.Prior.Inpainting.Harmonic.CNN}, the context encoder \cite{Marafioti2019:Context.encoder}, GACELA \cite{Marafioti2021:GACELA}, the diffusion-based MAID \cite{Liu2023:MAID}, and the current state-of-the-art,
\mbox{Janssen-TF}~\cite{MokryBalusikRajmic2025:Spectrogram.inpainting}.
However, the context encoder, GACELA, and MAID rely on offline pre-training on external datasets, and are moreover tailored to specific sampling rates different from that used in our evaluation. 
Comparing against them directly would not be fair, as performance differences would reflect dataset mismatches rather than algorithmic capability.
Therefore, our experimental comparison solely focuses on instance-optimized methods (DPAI, Janssen-TF, and \mbox{U-PHAIN-TF}) that operate on the spectrogram with no external training dependencies.
As an additional point of comparison, we also consider a~straightforward sparsity-based technique%
\footnote{The authors wish to thank one of the reviewers for this suggestion.
To the best of our knowledge, this method has not been previously published and thus can be considered a~minor contribution of this article.}
outlined below.

\subsection{Sparsity-based TF-inpainting ($\ell_1$-TF)}
\label{sec:L1.plain.TF.method}
Consider the optimization problem
\begin{equation}
    \label{eq:optimizationGCPA.L1}
    \argmin_{\x\in\Rset^L}
    \lambda\Vert\ana_{\g}\x\Vert_1 + \ind{\coefset}(\ana_{\g}\x).
\end{equation}
This problem exhibits the same structure as PHAIN-TF in \eqref{eq:optimizationGCPAphain} and can therefore be solved using the GCPA algorithm as well.
In this formulation, the phase correction step is omitted, such that directly the sparsity of the spectrogram is promoted.
This approach will be denoted $\ell_1$-TF in what follows.
As will become apparent later, the solution to this problem experiences energy loss in the gap, just like its time-domain counterpart.
The underlying cause is the same: the $\ell_1$ norm alone produces biased estimates
\cite{Chartrand2014:shrinkage.mappings,MokryRajmic2020:Inpainting.revisited}.

\subsection{Deep Prior Audio Inpainting (DPAI)}
DPAI is a deep-prior-based method that
uses a convolutional neural network (CNN) whose architecture itself (without external training) acts as the regularizer for audio inpainting.
The CNN is then overfitted using a single degraded spectrogram and a~solution is obtained by early stopping of the process.

In \cite{Miotello2023:Deep.Prior.Inpainting.Harmonic.CNN}, the authors consider a CNN described by the function $f_\theta(\Noise)$, where  \(\theta\) is a~set of trainable network parameters (weights and biases) and \(\Noise\in\Cset^{M \times N}\) is a random fixed noise matrix.
The main assumption is that the CNN can generate any spectrogram $\X\in\Cset^{M\times N}$ from $\Noise$ using a suitable $\theta$. 
Note that~\(\theta\) is initialized randomly and then optimized as
\begin{equation} \label{eq:implicitPrior}
    \theta^\star\! = \argmin_{\theta}E(\Tfmask\odot f_\theta(\Noise),\Xcor),
\end{equation}
where $E(\cdot)$ is a loss function that combines the mean squared error and the multi-scale spectrogram loss~\cite{Yamamoto2020:Parallel.waveGAN.multi-resolution.sgram}.
The optimal~$\theta^\star\!$ is acquired using standard optimizers such as Adam~\cite{Kingma2014:Adam}.
The reconstructed spectrogram $\hat{\X}$ is obtained by early stopping of the optimization, i.e., before overfitting occurs. Here, overfitting refers to the case where $f_{\theta^\star\!}(\Noise) = \Xcor$\!.
\subsection{Janssen-TF}
Janssen-TF is a recent adaptation of the Janssen inpainting algorithm~\cite{javevr86} to the TF case. 
It is based on autoregressive (AR) modeling. 
Thus, it utilizes harmonic components of a~spectrogram near the gaps to predict the values of missing TF coefficients.
It repeats two main steps in the $i$-th iteration.
First, given an estimate of a signal $\x^{(i)}$, compute the parameters of an AR model, which are denoted $\coef^{(i)}$\!.
Second, by creating a~Toeplitz matrix $\A^{(i)}$ composed of the coefficients $\coef^{(i)}$, estimate a new solution $\x^{(i+1)}$ as 
\begin{equation}
    \x^{(i+1)} = \arg\min_{\x} \tfrac{1}{2}\norm{\A\x}^2 + \ind{\coefset}(\ana_\g\x),
    \label{eq:subproblem}
\end{equation}
where $\coefset$ is the set \eqref{eq:setOfALLSpectrograms}.
The authors of \cite{MokryBalusikRajmic2025:Spectrogram.inpainting} used the alternating direction method of multipliers (ADMM) to solve the above problem, as it delivered the best performance.
However, other proximal algorithms are viable; see \cite{MokryBalusikRajmic2025:Spectrogram.inpainting}.

\section{Experiments and Results} \label{sec:expe}
\subsection{Datasets}
\label{sec:data}
Two datasets were chosen: the dataset used in \cite{Miotello2023:Deep.Prior.Inpainting.Harmonic.CNN, MokryBalusikRajmic2025:Spectrogram.inpainting},
hereafter referred to as the DPAI dataset, and the IRMAS dataset~\cite{Bosch2014:IRMAS.dataset}.

\subsubsection{DPAI dataset}
As mentioned above, DPAI will be utilized in the experiments as one of the reference methods.
The authors of DPAI tuned the network architecture and hyperparameters on their proprietary audio dataset, which is why these excerpts should be included for the sake of fair comparison.
The DPAI dataset consists of eight recordings, each 5 seconds long, sampled at 16\,kHz.

\subsubsection{IRMAS dataset}
The dataset comprises over 6000 audio recordings featuring various musical instruments, including cello, clarinet, flute, guitar, violin, and others, as well as recordings of a~singing voice.
Due to the large number of recordings, only a subset of IRMAS will be used. The subset\footnote{The list of the recordings used and the code to crop and subsample them can be found at \url{https://github.com/rajmic/spectrogram-inpainting/tree/main/audio-irmas}.}, also used in~\cite{MokryBalusikRajmic2025:Spectrogram.inpainting}, consists of 60 recordings, each 5 seconds long, sampled at 16\,kHz. 
With its 60 recordings, this dataset can provide more reliable results than the DPAI dataset in the objective comparison.

\subsection{Masks} \label{sec:gaps}
The same masks as in \cite{MokryBalusikRajmic2025:Spectrogram.inpainting} were chosen for fair comparison.
The masks are designed for 5-second spectrograms, containing one gap in each second.
The gap length ranges from 1 to 6 missing (zero) columns, 6 masks in total.
Figure \ref{fig:masks} illustrates three examples.
\begin{figure} [tb]
    \centering
    \adjustbox{width=.3\linewidth}{\input{tex/mask2}}%
    \hfill%
    \adjustbox{width=.3\linewidth}{\input{tex/mask4}}%
    \hfill%
    \adjustbox{width=.3\linewidth}{\input{tex/mask6}}
    \vspace{-2mm}
    \caption{Spectrogram masks used in experiments. Each mask has five gaps with lengths ranging from 1 missing column to 6 missing columns. Here, only the second mask (left), fourth mask (middle), and sixth mask (right) are shown.}
    \label{fig:masks}
\end{figure}
The relationship between the masked spectrogram columns and the corresponding affected time-domain samples was discussed in
Sec.~\ref{sec:T.and.TF.inpainting.nonequivalence}.
From this, we infer that masks 1, 2, and 3 do not erase samples completely, whereas masks 4, 5, and 6 produce trains of fully zeroed samples in the time domain, which in our setup amount to exactly 512, 1024, and 1536 zeroed samples, respectively (corresponding to 32, 64, and 96\,ms of silence).
In terms of the total number of signal samples affected by the spectrogram gaps, our configuration spans from 2048 samples (mask~1) up to 4608 samples (mask~6),
which is equivalent to a~duration of 128\,ms to 288\,ms.
\subsection{Metrics}
The Signal-to-Noise Ratio (SNR) is a widely recognized metric, defined in decibels as in~\cite{Adler2012:Audio.inpainting}:
 \begin{equation} \label{eq:SNRdefinition}
     \SNRfull(\xorig\!, \hat{\x}) =10\cdot\log_{10}\!\left(\dfrac{\norm{\xorig}^2}{\norm{\xorig-\hat{\x}}^2}\right),
 \end{equation}
where, in our case, $\hat{\x}$ is the
signal obtained from the reconstructed spectrogram by $\hat{\x}=\syn_\g\hat{\X}$. 

The Objective Difference Grade (ODG), computed using PEMO-Q v1.4.1 \cite{Huber:2006a}, is an estimate of subjective perception of restoration quality.  
It predicts the quality impairment of the reconstructed signal~\(\hat{\x}\) relative to the reference (original) signal \(\xorig\) on a continuous scale from \(-4\) to 0: Imperceptible~(0); Perceptible but not annoying ($-1$); Slightly annoying~($-2$); Annoying ($-3$); Very annoying ($-4$).

The ODG requires continuous signal context to assess perceptual quality and is therefore computed over the entire signal.
As for the SNR, certain studies on time-domain audio inpainting assess it only within the gaps ($\SNRgap$) \cite{Miotello2023:Deep.Prior.Inpainting.Harmonic.CNN, MokryRajmic2020:Inpainting.revisited,MokryZaviskaRajmicVesely2019:SPAIN,MokryRajmic2025:Inpainting.AR}, whereas other works \cite{TanakaYatabeOikawa2024:PHAIN}
compute it over the entire signal.
Nonetheless, assuming that the output of the method is guaranteed to lie within the feasible set $\sigset$,
it can be shown that $\SNRfull$ and $\SNRgap$ differ only by a factor constant for a~fixed signal.
Consequently, the differences in dB between methods exhibit exactly the same behavior for both the metrics.

Since our work is conducted in the TF domain, in addition to $\SNRfull$, we also report the SNR within the masked spectrogram regions, denoted by $\TFSNR$, which is computed in the same way as in \eqref{eq:SNRdefinition}.

\subsection{Implementation} \label{sec:implement}
The source codes for our proposed method, U-PHAIN-TF, are publicly available at \url{https://github.com/sedemto/phain-tf}. 
The proposed method was implemented in Matlab R2025a, drawing mainly on the Matlab codes\footnote{Matlab codes of U-PHAIN (and other types of PHAIN) are publicly available at \url{https://doi.org/10.24433/CO.1956970.v1}. 
In addition, certain functions from \url{https://doi.org/10.24433/CO.2743732.v1} were utilized.}~from the original \mbox{U-PHAIN}~\cite{TanakaYatabeOikawa2024:PHAIN}. 
In addition, we utilized the Large Time-Frequency Analysis Toolbox (LTFAT) \cite{LTFAT} for fast computation of time-frequency transformations%
\footnote{The codes for a faster computation of U-PHAIN are publicly available at \url{https://github.com/sedemto/phain-ltfat}.}
$\syn_\g$ and $\ana_\g$
required within \mbox{U-PHAIN}.

Before running the algorithm, the input data (corrupted spectrograms) are divided into smaller spectrogram segments positioned around the gaps. This is motivated by the reduction of the overall computational cost.
The context span is controlled by the hyperparameter \texttt{pad}, which specifies the minimum number of spectrogram columns taken from either side of the gap. 
The initial value of \texttt{pad} is the ratio of the effective window length to the hop size ($w/a$),
and the number of columns taken from each side of the gap is then adjusted to fulfill two conditions:
\begin{itemize}
    \item The STFT implementation in the LTFAT toolbox
    \cite{LTFAT}
    requires the segment length to be a~multiple of $M/a$.
    \item Since STFT with frequency-invariant phase is used, the starting index of the segment, minus one, must be divisible by $w/a$.%
    \footnote{In the case of the
    STFT with time-invariant phase, as employed in~\cite{MokryBalusikRajmic2025:Spectrogram.inpainting}, this particular condition can be disregarded.}
\end{itemize}
According to our findings \cite{Balusik2026acceleration}, the initial selection
$\texttt{pad} = w/a$ conforms to the recommended interval of $[w/a - 1, \, 2(w/a - 1)]$;
exceeding $2(w/a - 1)$ increases computational cost without enhancing reconstruction quality,
whereas lowering the value to $w/a - 1$ results in additional computational burden but with a~minor impact on the quality.
%

Peak values are computed from the associated time-domain signals and used to perform normalization of the segments.
The benefit of normalization is justified in Appendix~\ref{app:normal}.

After Algorithm~\ref{alg:gcpaUPHAINTF} has been executed on the normalized segments $\Xcor_\text{norm}$ and has produced the corresponding reconstructed spectrogram segments,
these segments are de-normalized and reintegrated into the originally observed corrupted spectrograms.


\begin{figure}
    \centering
    \adjustbox{width=.48\linewidth}{\input{tex/snr_iters}}%
    \hfill%
    \adjustbox{width=.51\linewidth}{\input{tex/odg_iters}}%
    \vspace{-2mm}%
	\caption{%
        Results from U-PHAIN-TF with different settings of inner iterations, while the number of outer iterations was fixed to $J=10$ and $\lambda=0.01$.
        The $\SNRfull$ (left) and ODG (right) are computed on the DPAI dataset and averaged across all examples and masks.
	}%
    \label{fig:itersInner}
\end{figure}

\begin{figure}
    \centering
    \adjustbox{width=.48\linewidth}{\input{tex/outerIters-SNR}}%
    \hfill%
    \adjustbox{width=.5\linewidth}{\input{tex/outerIters-ODG}}%
    \vspace{-2mm}%
	\caption{%
        Results from U-PHAIN-TF with different settings of outer iterations, while the number of inner iterations was fixed to $I=500$ and $\lambda=0.01$.
        The $\SNRfull$ (left) and ODG (right) are computed on the DPAI dataset and averaged across all examples and masks.
	}%
    \label{fig:outerInner}
\end{figure}

\subsection{Choice of Hyperparameters}
\label{sec:hyperparameters.choice}
U-PHAIN-TF involves a number of hyperparameters,
which include the STFT setting (window length $w$, window type, hop size $a$, FFT length $M$),
parameters of the GCPA itself
(step sizes $\sigma$, $\eta$, $\tau$, threshold~$\varepsilon$),
as well as the number of inner $I$ and outer iterations $J$, and parameters \texttt{warmRestart},
and \texttt{pad}.
In addition, a relative flooring factor $\beta$ is utilized for numerical stability \cite{Yatabe2019:representation.complex.spec}.
Finally, $\lambda$ appears as a regularization parameter in the optimization problem \eqref{eq:optimizationGCPAphain} being solved,
but in our context it acts more as a parameter of the GCPA algorithm itself,
as discussed in Sec.\,\ref{sec:uphaintf}.

\subsubsection{STFT parameters}
For the STFT, we use a 2048-sample-long Hann window (corresponding to 128\,ms) with a 75\% overlap and 2048 frequency channels, such that $\ana_\g$ forms a~Parseval tight frame.
The parameters are the same as in~\cite{MokryBalusikRajmic2025:Spectrogram.inpainting} and they were chosen to provide a~fair comparison with the reference methods.
In addition, we calculate the time-directional derivative of the window
(i.e., $\g'$ required in~\eqref{eq:instantPHAIN}) analytically. 
Alternatively, a spectral derivative may be employed; see \cite{Kusano2022:windowFunctionsSpectralDerivative}.

\subsubsection{Step sizes} The step sizes for U-PHAIN-TF are chosen such that
$\tau\cdot\sigma\cdot\norm{\D\R_{\insF}\ana_{\g}}^2\leq1$
and $\tau\cdot\eta\cdot\norm{\ana_\g}^2\leq1$.
From Section~\ref{sec:stft}, it follows that  \(\norm{\ana_\g}^2=1\); thus, \(\eta  = 4\) and \(\tau = 0.25\) to satisfy the upper limit of the condition.
In addition, with the help of a~shift operator $\mathcal{A}$ that translates any spectrogram~$\X$ along the time axis, i.e.,\ $(\mathcal{A}\X)[m,n] = X[m,n+1]$, the operator norm of $\D\R_\insF$ can be upper-bounded:
\begin{equation}\label{eq:DRoperator}
    \norm{\D\R_\insF}= \norm{ \R_\insF -\mathcal{A} \R_\insF }\leq\norm{\R_\insF}+\norm{\mathcal{A}\R_\insF}= 2. 
\end{equation}
This relation is possible thanks to the fact that
$\R_\insF$ is unitary, \(\norm{\R_\insF}=1\),
and $\mathcal{A}$ is bounded such that \(\norm{\mathcal{A}}=1\)~\cite{bayram2013:simple_prior}.
Using the above, the operator norm is bounded such that $\norm{\D\R_{\insF}\ana_{\g}}^2\leq4$.
Although this bound is not necessarily tight, the step sizes \(\sigma  = 1\) and \(\tau = 0.25\) are a valid choice as they satisfy the convergence condition at the upper limit.
Consequently, weak convergence
is guaranteed within each inner loop.

\subsubsection{Iterations}
We have also investigated an appropriate number of inner ($I$) and outer iterations ($J$) in the proposed context of spectrogram inpainting.
First, we fixed $J$ to 10 (same as in \cite{TanakaYatabeOikawa2024:PHAIN}) and set the stopping threshold $\varepsilon=0.001$.
Then, we computed the metrics on the DPAI dataset while changing~$I$ (i.e., the iteration count of the GCPA).
The results of this experiment are shown in Fig.~\ref{fig:itersInner}.
We observe that the metrics start to change more slowly at 
around 500 iterations.
Thus, we selected $I=500$ as it offers high reconstruction quality while saving some computational time.
Finally, we examined the parameter $J$, i.e., the number of instantaneous frequency updates.
The experiments shown in Fig.~\ref{fig:outerInner} indicate that the setting $J=10$ is sufficient and that larger values of $J$ yield little or no additional benefit.
In many cases, the stopping criterion $\norm{\hat{\x}^{(j)}-\hat{\x}^{(j-1)}}_2<\varepsilon$ was even met before reaching $J=10$.

In the algorithm, two additional parameters are utilized:
The flooring factor $\beta$ is set to $\beta=10^{-10}$ to prevent division by zero, following \cite{Yatabe2019:representation.complex.spec}.
We set \texttt{warmRestart = false} for our experiments; however, setting it to \texttt{true} can yield additional improvement; see the accompanying webpage\footnote{\url{https://sedemto.github.io/phain-tf/}\label{fn:webpage}}.

\begin{figure} 
    \centering
    \adjustbox{width=.8\linewidth}{\input{tex/lambda_gcpa}}%
    \vspace{-2mm}%
	\caption{%
        Objective evaluation for U-PHAIN-TF with $I=500$, for different settings of $\lambda$. The ODG results are computed on both datasets, with all masks applied to each audio example. For each \(\lambda\), the results are averaged.
	}%
    \label{fig:lambdas}
\end{figure}
\subsubsection{Lambda}
\label{sec:hyperparameters.choice.lambda}
Given that the input data are normalized (see Appendix \ref{app:normal}), the optimal value of $\lambda$ can be determined.
To find it, we computed the ODG metric on the reconstructions from both the DPAI and IRMAS datasets, for $\lambda$ values ranging from $10^{-7}$\! to $10^2$\!.
As shown in Fig.~\ref{fig:lambdas}, the optimal $\lambda$ value is approximately~$10^{-2}$\! for both datasets, indicating that such a~hyperparameter choice is dataset-independent and transferable.
Note that this observation likewise holds when SNR is considered.
Finally, note that the optimal value of $\lambda$ may change when the number of inner iterations is varied.

\subsection{Expectations}
\label{sec:expectations.about.results}
The STFT is redundant because multiple time-shifted windows overlap and include a~time-domain sample.
In our specific setup, each time-domain sample affects four spectrogram columns.
Since our STFT forms a~tight frame, the analysis and synthesis window systems coincide
(as in \eqref{eq:STFTdefnition} and \eqref{eq:STFTdefnition.synthesis}),
so in principle up to three spectrogram columns can be dropped out and still allow for uniquely recovering the signal.
This can be done by a~straightforward compensation of the corresponding energy loss.
As a~consequence, reconstructions for masks 1, 2, and 3
(i.e., one to three missing columns) are expected to be identical and essentially independent of the chosen regularizer.
In practice, as shown in the following sections, this expectation holds for mask~1 and mask~2, whereas for mask~3 the performance degrades rapidly because the associated energy curve can touch zero, leading to severe numerical instabilities.
(Note that even a no-prior setting (Alg.~\ref{alg:gcpaUPHAINTF} with $\lambda=0$) yields near-perfect reconstructions for masks~1 and 2, as illustrated on the accompanying webpage\footref{fn:webpage}.)%

\begin{figure}
    \centering
    \adjustbox{width=.51\linewidth}{\input{tex/snr_in_gaps_dpai}}%
    \adjustbox{width=.5\linewidth}{\input{tex/odg_dpai_withL1}}%
    \vspace{-2mm}%
	\caption{%
        Comparison on the DPAI dataset of the proposed method with other TF domain inpainting methods in terms of $\TFSNR$ (left) and ODG (right). 
        Wherever the confidence intervals do not overlap, it
        may be concluded that the difference of the means is statistically significant.
	}%
    \label{fig:dpaiResults}
\end{figure}

\subsection{Objective Evaluation}
\label{sec:objectiveEval}
This section compares the proposed method with other TF-domain inpainting methods, specifically DPAI \cite{Miotello2023:Deep.Prior.Inpainting.Harmonic.CNN}, Janssen-TF \cite{MokryBalusikRajmic2025:Spectrogram.inpainting}, and the sparsity-based reference, $\ell_1$-TF, briefly described in Sec.~\ref{sec:L1.plain.TF.method}.
For DPAI, the ``DPAI with context'' variant using the ``best2'' architecture was employed, which was its strongest-performing configuration in~\cite{MokryBalusikRajmic2025:Spectrogram.inpainting}.
For \mbox{$\ell_1$-TF}, 500 iterations of the GCPA were run, and
the optimal $\lambda$ value of $10^{-3}$ was determined as in
Sec.~\ref{sec:hyperparameters.choice}. 
The results were averaged over all the examples in each dataset for every mask (see Section \ref{sec:gaps}).
In the figures that follow, the horizontal axis represents the masks as ``gap length'', while
the vertical axis represents the corresponding objective metric (ODG, $\SNRfull$, $\TFSNR$).
The execution time was also measured for all methods.

\subsubsection{Results on DPAI dataset}
The objective metrics computed on the DPAI dataset are shown in Fig.~\ref{fig:dpaiResults}.
The bold central lines show the mean values, while the shaded regions around them denote the 95\% confidence intervals for the means. 
Since the DPAI dataset contains only eight examples, the confidence intervals are relatively wide.
Nonetheless, on this small dataset, \mbox{U-PHAIN-TF} achieved the best performance in terms of ODG,
but the differences are not statistically significant for most of the gap sizes, compared with Janssen-TF and \mbox{$\ell_1$-TF}.
The baseline, $\ell_1$-TF, achieved ODG scores comparable to \mbox{U-PHAIN-TF} up to the mask size three, after which its performance dropped rapidly.
As for the SNR, \mbox{U-PHAIN-TF} achieved results comparable (large gaps) or greater (short gaps) to those of \mbox{Janssen-TF}.
Compared with \mbox{$\ell_1$-TF}, \mbox{U-PHAIN-TF} yielded similar or slightly worse SNR values for shorter gaps; however, its advantages become evident for larger gaps.
Due to its simplicity, the $\ell_1$-TF is particularly prone to energy loss, especially in cases of masks 4--6 that lead to zero intervals in the time domain.
By incorporating iPCTV minimization, U-PHAIN-TF addressed this energy loss problem, resulting in a higher overall reconstruction quality.

In Fig.~\ref{fig:dpaiResults}, note the exceptionally high scores for masks 1 and 2.
This observation is consistent with the redundancy of the STFT, as discussed in Sec.~\ref{sec:expectations.about.results}.
The differences in the attained values for these masks can be attributed to the use of distinct reconstruction algorithms with different convergence behavior.

%

\begin{figure}

    \centering
    \adjustbox{width=.510\linewidth}{\input{tex/snr_in_gaps_irmas}}%
    \adjustbox{width=.500\linewidth}{\input{tex/odg_irmas_withL1}}%
    \vspace{-2mm}%
	\caption{%
        Comparison on the IRMAS dataset
        with other TF domain inpainting methods
        in terms of mean $\TFSNR$ (left) and ODG (right).
	}%
    \label{fig:irmasResults}
\end{figure}
\subsubsection{Results on IRMAS dataset}
Our IRMAS subset with its 60 excerpts allows us to obtain statistically more significant results.
The objective metrics
are shown in Fig.~\ref{fig:irmasResults}:
in terms of ODG, U-PHAIN-TF clearly surpassed all other inpainting methods.
The confidence interval of \mbox{U-PHAIN-TF} does not overlap with those of DPAI and \mbox{Janssen-TF}.
Meanwhile, there is an overlap of $\ell_1$-TF with that of U-PHAIN-TF only for the first two masks.
The SNR results are similar to those obtained on the smaller DPAI dataset:
for shorter gaps, U-PHAIN-TF outperformed Janssen-TF and yielded results comparable to $\ell_1$-TF.
For larger gap lengths, U-PHAIN-TF and Janssen-TF achieved comparable SNR results,
outperforming both $\ell_1$-TF and DPAI by a~statistically significant margin.

Figure \ref{fig:scatter} shows additional scatter plots that compare the objective scores of Janssen-TF with those of U-PHAIN-TF in more detail. 
The SNR values show that for larger gaps, the methods perform similarly, confirming the previous cumulative findings. 
However, the ODG results are strongly biased towards
\mbox{U-PHAIN-TF}.

\begin{figure}
    \centering
    \adjustbox{width=.36\linewidth}{\input{tex/snrScatter1}}%
    \adjustbox{width=.32\linewidth}{\input{tex/snrScatter2}}%
    \adjustbox{width=.32\linewidth}{\input{tex/snrScatter3}}%
    
    \adjustbox{width=.36\linewidth}{\input{tex/snrScatter4}}%
    \adjustbox{width=.32\linewidth}{\input{tex/snrScatter5}}%
    \adjustbox{width=.32\linewidth}{\input{tex/snrScatter6}}%

    \adjustbox{width=.36\linewidth}{\input{tex/odgScatter1}}%
    \adjustbox{width=.33\linewidth}{\input{tex/odgScatter2}}%
    \adjustbox{width=.31\linewidth}{\input{tex/odgScatter3}}%
    
    \adjustbox{width=.35\linewidth}{\input{tex/odgScatter4}}%
    \adjustbox{width=.32\linewidth}{\input{tex/odgScatter5}}%
    \adjustbox{width=.33\linewidth}{\input{tex/odgScatter6}}%
    \vspace{-1mm}%
	\caption{%
       Scatter plots comparing $\SNRfull$ and ODG scores of Janssen-TF and U-PHAIN-TF for each mask (1--6 missing spectrogram columns). The gray line indicates identical performance.
	}%
    \label{fig:scatter}
\end{figure}
\begin{table*} 
    \centering
    \caption{Computational demand of involved TF-inpainting methods.
    Processing times were averaged across examples in the DPAI dataset.}
    \renewcommand{\arraystretch}{1.1}
    \begin{tabular}{|c|c|r|r|r|r|r|r|}
        \hline
        \multirow{2}{*}{\textbf{Method}} &  \multirow{2}{*}{\textbf{CPU/GPU; RAM/VRAM}} & \multicolumn{6}{c|}{\textbf{Execution Time [s]}} \\
        \cline{3-8}
        & & \textbf{mask\,1} & \textbf{mask\,2} & \textbf{mask\,3} & \textbf{mask\,4} & \textbf{mask\,5} & \textbf{mask\,6} \\ \hline\hline
         DPAI \cite{Miotello2023:Deep.Prior.Inpainting.Harmonic.CNN} &  NVIDIA Tesla V100S; 32\,GB &  1\,140.0& 1\,140.0 & 1\,140.0 & 1\,140.0 & 1\,140.0 & 1\,140.0 \\ \hline
         Janssen-TF \cite{MokryBalusikRajmic2025:Spectrogram.inpainting}& \multirow{4}{*}{AMD Ryzen 9 9900X, 4.4\,GHz; 64\,GB} &  123.2 &123.4&202.8&202.9&202.9&205.2 \\ \cline{1-1}\cline{3-8}
         $\ell_1$-TF&  & 1.9 &1.9&2.0&2.1&2.1& 2.3 \\ \cline{1-1}\cline{3-8}
         U-PHAIN-TF&  & 5.5&5.7&29.1&30.5&31.3&34.2  \\ \cline{1-1}\cline{3-8}
         U-PHAIN-TF (no segmentation)&  & 10.9&13.0&54.6&54.4&54.1&54.1  \\ \hline
    \end{tabular}
    \label{tab:execTime}
\end{table*}

\subsubsection{Execution time}
In our previous study
\cite{MokryBalusikRajmic2025:Spectrogram.inpainting},
we have reported
that reconstructing a 5-second spectrogram
using the DPAI method takes approximately 19 minutes on an NVIDIA Tesla V100S GPU with 32\,GB of memory, regardless of the number of gaps and their length.
We run the same reconstruction task with Janssen-TF, $\ell_1$-TF and U-PHAIN-TF
(in Matlab R2025a);
the cost of these methods increases with the gap length.
The timing is summarized in Table \ref{tab:execTime}.
The execution time of U-PHAIN-TF is demonstrably the lowest among all methods, taking only about 6–34 seconds.

\begin{figure}[tb]
    \vspace{4pt}
    \centering
    \adjustbox{width=.95\linewidth}{\input{tex/listeningNew}}
    \vspace{-1ex}%
	\caption{%
        A boxplot showing the distribution of scores in the listening test.
        The individual boxes span from the 25th to the 75th percentile. 
        The notches (filled areas) around the medians are the 95\% confidence intervals.
	}%
    \label{fig:listening}
\end{figure}

\subsection{Subjective Evaluation}
To confirm 
the objective comparison, a listening test was conducted using the MUSHRA method \cite{ITU-R2015:MUSHRA}.
To fulfill the requirements of MUSHRA, i.e., to prevent listener fatigue, we evaluated a~subset of conditions:
three gap lengths (2, 4, and~6),
across six audio examples from the DPAI dataset (0, 1, 3, 4, 5, and 7).
Examples 2 and 6 were excluded \textit{a~priori} due to high background noise and irregular temporal structures.
U-PHAIN-TF was evaluated against DPAI and Janssen-TF methods. 
Audio excerpts used in the experiment, additional supporting material, and reconstructions from the \mbox{$\ell_1$-TF} baseline are available online at the accompanying webpage\footref{fn:webpage}.
Despite the comparatively high ODG scores, the \mbox{$\ell_1$-TF} reconstructions exhibit clearly audible energy-loss artifacts (volume pumping), particularly for larger gaps.

For the test, the webMUSHRA environment was utilized~\cite{Schoeffler2018:webMUSHRA}.
There were five conditions in total: a hidden reference signal, an anchor signal (the corrupted signal computed from the corrupted spectrogram), and three reconstructions corresponding to each of the inpainting methods. 
With six examples and three masks, the entire test consisted of 18 test pages.
Participants rated how similar each signal was to the reference on a continuous scale from 0 to 100.
The test took place in a quiet music studio using the PreSonus AudioBox USB 96 sound card and Sennheiser HD 700 headphones.
The listening conditions were identical for all participants.
In line with \cite{ITU-R2015:MUSHRA}, 17 of the 21 assessors were retained after post-screening.

The results
are depicted in a~box plot in Fig.~\ref{fig:listening}.
\mbox{U-PHAIN-TF} achieves the highest median score across all methods,
suggesting it is the most promising option for time-frequency domain inpainting.
The superiority of U-PHAIN-TF
is supported by the Wilcoxon signed-rank test\footnote{The results of this test are available at the accompanying webpage.}, which rejects the null hypothesis that the medians of the \mbox{U-PHAIN-TF} and Janssen-TF score distributions are the same, in favor of the alternative that \mbox{U-PHAIN-TF} has a higher median.
(In the same sense, Janssen-TF exhibits a~higher median than DPAI.)

The box plot of subjective scores \emph{per gap}, provided on the accompanying webpage\footref{fn:webpage}, corroborates the finding that for mask~2, \mbox{U-PHAIN-TF} and Janssen-TF deliver quality that cannot be distinguished from the reference. 

\section{Conclusion} \label{sec:conclude}
In this paper, we proposed a novel method for spectrogram inpainting, U-PHAIN-TF,
and compared it to existing
methods
(DPAI, Janssen-TF, $\ell_1$-TF).
Thanks to the incorporation of the phase-aware prior \cite{YatabeOikawa2018:ipctv,TanakaYatabeOikawa2024:PHAIN},
\mbox{U-PHAIN-TF} alleviates the energy loss typically observed in sparsity-based approaches.
In terms of SNR, the performance of \mbox{U-PHAIN-TF} is on par with its main competitor, \mbox{Janssen-TF}.
With respect to perceptual quality (both objective and subjective), however, \mbox{U-PHAIN-TF} consistently outperformed all reference methods.
In addition to its qualitative performance, U-PHAIN-TF is considerably less computationally demanding than DPAI and Janssen-TF.
\appendices
\section{An Alternative Approach} \label{app:alternative}
The problem \eqref{eq:optimizationGCPAphain} is not the only possible formulation for spectrogram inpainting that involves a~phase-aware prior.
The problem can also be written as seeking a~solution directly in the time-frequency domain:
\begin{equation}
    \label{eq:formulation.in.TF}
    \argmin_{\X\in\Cset^{M \times N}}\lambda\iPCTVtf+\ind{\coefset}(\X).
\end{equation}
In contrast to \eqref{eq:optimizationGCPAphain},
the standard CPA \cite{Condat2023:Proximal.splitting.algorithms}
is sufficient to solve \eqref{eq:formulation.in.TF}.
The hyperparameters were set as follows: $\lambda=10$, $I=500$, and $J=10$.
Within the resulting iterative scheme, neither forward nor inverse transformations between the time and time-frequency domains are required,
so that both the primal and dual variables reside
exclusively in the time-frequency domain.
If needed for subsequent evaluation, the spectrogram can be transformed into the time domain as a~final post-processing step.

The results comparing the proposed method with this alternative, computed on the \mbox{IRMAS} dataset, are shown in Fig.~\ref{fig:alternativeMethod}.
The alternative method performs significantly worse in both SNR and ODG.
Results suggest that alternating between the time and time-frequency domains during the iterative process is beneficial, and perhaps even essential.
Such an observation can be supported by theoretical arguments.
In Alg.~\ref{alg:gcpaUPHAINTF}, each (inner) iteration contains application of both $\ana_\g$ and $\syn_\g$.
Specifically, its lines
\ref{alg:line:Yupdate} to \ref{alg:line:Qupdate}
can be interpreted as updates of variables in the TF domain,
and this update involves the composite operator $\ana_\g\syn_\g$.
This operator can be interpreted as a projection of a spectrogram onto the range space of $\ana_\g$.
According to Sec.~\ref{sec:stft}, this space contains so-called consistent spectrograms.
Consequently, each iteration adjusts the current TF variables toward the space of consistent spectrograms.
By contrast, the operators $\ana_\g$ and $\syn_\g$ are entirely absent from the iterations of the algorithm that solves \eqref{eq:formulation.in.TF}, which keeps the solution far from that space and thereby degrades the reconstruction quality.
To further emphasize the value of moving between the two domains, note that \eqref{eq:optimizationGCPA.L1} is actually a simpler formulation than~\eqref{eq:formulation.in.TF}.
Nevertheless, because it is solved in the time domain, 
the operator $\ana_\g\syn_\g$ is present in the algorithm, which leads to substantially better performance than \eqref{eq:formulation.in.TF}.


\begin{figure}
    \centering
    \adjustbox{width=.505\linewidth}{\input{tex/irmas_snr_alt}}%
    \adjustbox{width=.495\linewidth}{\input{tex/irmas_odg_alt}}%
    \vspace{-2mm}%
	\caption{%
        Comparison on the IRMAS dataset of the proposed method with its alternative version solving
        \eqref{eq:formulation.in.TF}
        that resides solely in the TF domain.
        The average SNR (left) and ODG (right) are shown, with confidence intervals.
	}%
    \label{fig:alternativeMethod}
\end{figure}


\section{Thresholding and Norms} \label{app:thresholding}
Similar to \cite{TanakaYatabeOikawa2024:PHAIN}, we investigated the performance of the proposed method (\mbox{U-PHAIN-TF}) when the soft thresholding in Alg. \ref{alg:gcpaUPHAINTF} is replaced by its alternatives such as $p$-shrinkage
or smooth-hard thresholding~\cite{Chartrand2014:shrinkage.mappings}.
The \mbox{$p$-shrinkage} operator is defined 
\cite{Chartrand2014:shrinkage.mappings}
\begin{equation*}
    \mathrm{shrink}_p(\X) = \sgn(\X)\odot \max (|\X|-\lambda^{2-p}|\X|^{p-1},0),
\end{equation*}
where $\X$ is a spectrogram and operations are performed elementwise;
$\lambda$ is a~threshold below which elements are set to zero,
as in \eqref{eq:softThresholding}.
For $p=1$, this operator coincides with the soft thresholding \eqref{eq:softThresholding}.
Similarly, the smooth-hard thresholding operator is defined as
\begin{equation*}
    \mathrm{SH}(\X) = \X\eul^{-\alpha/(\eul^{|\X|-\lambda}-1)^2}\ \text{for}\ |\X| \geq \lambda,\ \text{otherwise 0}.
\end{equation*} 
As in \cite{Chartrand2014:shrinkage.mappings}, $\alpha$ is an extra tuning parameter.
We then performed several experiments to determine the optimal $\lambda$ for each thresholding case.
We evaluated $\lambda$ values from $10^{-5}$ to $10^{2}$ and selected hyperparameters as follows:
For $p$-shrinkage, we chose $p =0.9$ paired with $\lambda=0.01$.
We explored values of $p$ in $(-1,1)$.
We observed that performance improves as $p$ approaches the soft-thresholding case.
For smooth-hard thresholding, we set $\alpha=10^{-2}$ as in \cite{Chartrand2014:shrinkage.mappings}, and $\lambda =10^{-3}$\!.
For the shrinkage operator corresponding to the $\ell_2$-norm,
we have found that the optimal $\lambda$ was $1$.
For $\ell_2$-squared variant
(denoted $\ell_2^2$ in Fig.~\ref{fig:thresholding}),
the best performing $\lambda$ was $0.2$.

We compared all of the above cases with \mbox{U-PHAIN-TF} and its basic variant (denoted \mbox{B-PHAIN-TF}), i.e., the case omitting the IF update, in terms of both SNR and ODG.
In addition, we included an oracle version of \mbox{B-PHAIN-TF}, where the IF is calculated from the ground truth spectrogram (unknown in practice). 
The ODG results of this experiment are shown in Fig.~\ref{fig:thresholding}.
We observe that \mbox{U-PHAIN-TF} performs the best when using the \mbox{$\ell_1$-norm}.
This choice thus works well both in the original time-domain method of
\cite{TanakaYatabeOikawa2024:PHAIN}
and in its proposed time-frequency adaptation.

\begin{figure}
    \centering
    \adjustbox{width=.9\linewidth}{\input{tex/thresholding}}%
    \vspace{-2mm}%
	\caption{%
        Comparison on the DPAI dataset of U-PHAIN-TF with different thresholding operators. Only the mean ODG value is shown.
	}%
    \label{fig:thresholding}
\end{figure}
\section{Normalization} \label{app:normal}
In the paper, we mention normalization of the input data for Alg.~\ref{alg:gcpaUPHAINTF}
(i.e., scaling of corrupted spectrogram segments).
To assess whether normalization is necessary, we run \mbox{U-PHAIN-TF} under two conditions: with and without normalization, and compare the reconstructions. 
If the objective results differ, i.e., the optimal lambda value changes in each test, normalization is required.
We adopt the following procedure:
\begin{enumerate}
    \item Define a (logarithmically spaced) set of ten values of  $\lambda$: \(\lambda=\{10^{-7}\!,10^{-6}\!,\dots,10^{1}\!,10^{2}\}\).
    \item For each $\lambda$, run \mbox{U-PHAIN-TF} on the DPAI dataset and compute the mean ODG results.
    \item Repeat the previous step; this time normalize the amplitude of each input corrupted spectrogram segment.
\end{enumerate}
The mean ODG values are shown in Fig.~\ref{fig:normalization}.
The results show that the optimal lambda differs for each test, demonstrating the need for input data normalization.
Without normalization, the optimal lambda value would vary for each input.

\begin{figure}
    \centering
    \adjustbox{width=.8\linewidth}{\input{tex/normalization}}%
    \vspace{-2mm}%
	\caption{%
        The normalization test on U-PHAIN-TF. For each amplitude test and lambda value, the ODG metric is calculated for all reconstructions. Only the mean ODG value is shown.
	}%
    \label{fig:normalization}
\end{figure}


\bibliographystyle{IEEEtran}
\
\newcommand{\noopsort}[1]{} \newcommand{\printfirst}[2]{#1} \newcommand{\singleletter}[1]{#1} \newcommand{\switchargs}[2]{#2#1}

\newpage

\end{document}

%% file: defs.tex
\DeclareMathOperator*{\argmin}{argmin}

\newcommand{\qm}[1]{``#1''}

\newcommand{\x}{\mathbf{x}} 
\newcommand{\xorig}{\mathbf{x}^{\textup{orig}}} 
\newcommand{\Noise}{\mathbf{N}} 
\newcommand{\coef}{\mathbf{a}} 
\newcommand{\mask}{\mathbf{m}} 
\newcommand{\xcor}{\mathbf{x}^{\textup{cor}}} 
 
\newcommand{\g}{\mathbf{g}} 
\newcommand{\insF}{{\boldsymbol{\omega}}}
\newcommand{\insFreq}[1]{\boldsymbol{\omega}_{\mathbf{#1}}} %

\newcommand{\X}{\mathbf{X}} 
\newcommand{\Y}{\mathbf{Y}}
\newcommand{\Z}{\mathbf{Z}}
\newcommand{\sQ}{\mathbf{Q}}
\newcommand{\sR}{\mathbf{R}}
\newcommand{\Xorig}{\mathbf{X}^{\textup{orig}}} 
\newcommand{\Xcor}{\mathbf{X}^{\textup{cor}}} 
\newcommand{\Tfmask}{\mathbf{M}} 
\newcommand{\A}{\mathbf{A}} 
\newcommand{\ana}{\mathcal{G}} 
\newcommand{\syn}{\mathcal{G}^*} 
\newcommand{\R}{\mathcal{R}} 
\newcommand{\Rad}{\mathcal{R}^*} 
\newcommand{\D}{\mathcal{D}} 
\newcommand{\Dad}{\mathcal{D}^*} 
\newcommand{\Id}{\mathit{Id}} 

\newcommand{\Cset}{\mathbb{C}} 
\newcommand{\Rset}{\mathbb{R}} 
\newcommand{\Nset}{\mathbb{N}} 
\newcommand{\ind}[1]{\iota_{#1}} 
\newcommand{\sigset}{\mathit{\Gamma}_{\textup{T}}} 
\newcommand{\coefset}{\mathit{\Gamma}_{\textup{TF}}} 

\newcommand{\jmag}{\mathrm{i}}
\newcommand{\eul}{\mathrm{e}}

\newcommand{\norm}[1]{\|#1\|}
\newcommand{\normOne}[1]{{\Vert#1\Vert}_1}

\newcommand{\proj}[1]{\operatorname{proj}_{#1}}
\newcommand{\prox}[1]{\operatorname{prox}_{#1}}
\newcommand{\soft}[1]{\operatorname{soft}_{#1}}
\newcommand{\inner}[2]{{\langle#1,#2\rangle}}

\newcommand{\iPCTVwhole}{\normOne{\D\R_{\insF}\ana_{\g}\x}}
\newcommand{\iPCTVtf}{\normOne{\D\R_{\insF}\X}}

\newcommand{\SNRgap}{\mathrm{SNR}_\mathrm{gap}}
\newcommand{\SNRfull}{\mathrm{SNR}_{\mathrm{full}}}
\newcommand{\TFSNR}{\mathrm{TF\text{-}SNR}_{\mathrm{gap}}}
\DeclareMathOperator{\sgn}{sgn}

%% file: tex/mask2.tex
%
%
\begin{tikzpicture}

\begin{axis}[%
width=0.45\linewidth,
height=0.4\linewidth,
at={(0in,0in)},
scale only axis,
axis on top,
xmin=0,
xmax=5,
xtick={0, 1, 2, 3, 4, 5},
xlabel={time [s]},
y dir=reverse,
ymin=0,
ymax=1025,
ytick={\empty},
axis background/.style={fill=white}
]
\addplot [forget plot] graphics [xmin=0, xmax=5, ymin=0, ymax=1025] {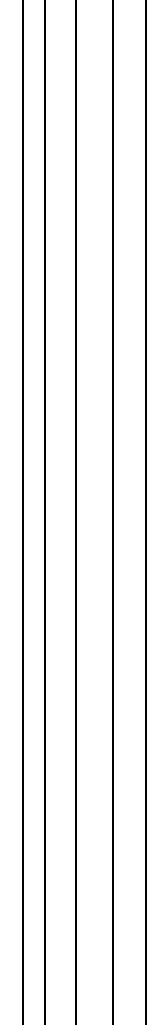};
\end{axis}
\end{tikzpicture}%

%% file: tex/mask4.tex
%
%
\begin{tikzpicture}

\begin{axis}[%
width=0.45\linewidth,
height=0.4\linewidth,
at={(0in,0in)},
scale only axis,
axis on top,
xmin=0,
xmax=5,
xtick={0, 1, 2, 3, 4, 5},
xlabel={time [s]},
y dir=reverse,
ymin=0,
ymax=1025,
ytick={\empty},
axis background/.style={fill=white}
]
\addplot [forget plot] graphics [xmin=0, xmax=5, ymin=0, ymax=1025] {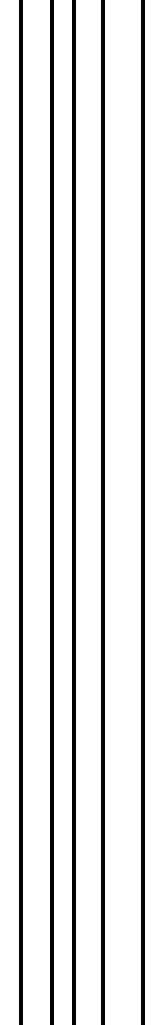};
\end{axis}
\end{tikzpicture}%

%% file: tex/mask6.tex
%
%
\begin{tikzpicture}

\begin{axis}[%
width=0.45\linewidth,
height=0.4\linewidth,
at={(0in,0in)},
scale only axis,
axis on top,
xmin=0,
xmax=5,
xtick={0, 1, 2, 3, 4, 5},
xlabel={time [s]},
y dir=reverse,
ymin=0,
ymax=1025,
ytick={\empty},
axis background/.style={fill=white}
]
\addplot [forget plot] graphics [xmin=0, xmax=5, ymin=0, ymax=1025] {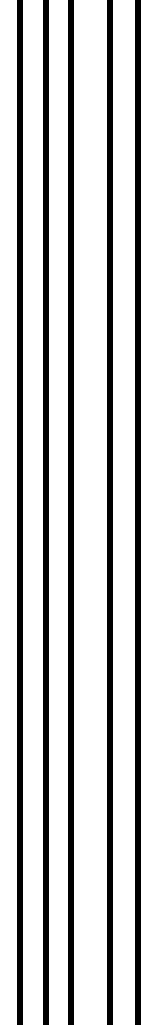};
\end{axis}
\end{tikzpicture}%

%% file: tex/snr_iters.tex
%
%
\definecolor{mycolor1}{rgb}{0.00000,0.44700,0.74100}%
\begin{tikzpicture}

\begin{axis}[%
width=0.45\linewidth,
height=0.4\linewidth,
at={(0in,0in)},
scale only axis,
xmin=0,
xmax=1000,
xlabel style={font=\color{white!15!black}},
xlabel={number of inner iterations},
ymin=30,
ymax=55,
ylabel style={font=\color{white!15!black}},
ylabel={$\SNRfull$ [dB]},
axis background/.style={fill=white},
xmajorgrids,
ymajorgrids
]
\addplot [color=mycolor1, line width=1.0pt, mark=x, mark size=4pt, mark options={solid, mycolor1}, forget plot]
  table[row sep=crcr]{%
50	32.1119779977912\\
100	36.7711316648364\\
200	42.1312256245118\\
300	46.5967374311341\\
500	50.9409083491181\\
800	53.179714807386\\
1000	54.1379190251681\\
};
\end{axis}

\end{tikzpicture}%

%% file: tex/odg_iters.tex
%
%
\definecolor{mycolor1}{rgb}{0.00000,0.44700,0.74100}%
\begin{tikzpicture}

\begin{axis}[%
width=0.45\linewidth,
height=0.4\linewidth,
at={(0in,0in)},
scale only axis,
xmin=0,
xmax=1000,
xlabel style={font=\color{white!15!black}},
xlabel={number of inner iterations},
ymin=-1.5,
ymax=-0.9,
ylabel style={font=\color{white!15!black}},
ylabel={ODG},
axis background/.style={fill=white},
xmajorgrids,
ymajorgrids
]
\addplot [color=mycolor1, line width=1pt, mark=x, mark size=4pt, mark options={solid, mycolor1}, forget plot]
  table[row sep=crcr]{%
50 -1.41040893428693\\
50 -1.41040893428693\\
100	-1.24874869219547\\
200	-1.07932399028098\\
300	-1.02249896112922\\
500	-0.964841404943428\\
800	-0.925026021599533\\
1000 -0.917860666042241\\
};
\end{axis}

\end{tikzpicture}%

%% file: tex/outerIters-SNR.tex
%
%
\definecolor{mycolor1}{rgb}{0.06600,0.44300,0.74500}%
\definecolor{mycolor2}{rgb}{0.12941,0.12941,0.12941}%
\begin{tikzpicture}

\begin{axis}[%
width=0.45\linewidth,
height=0.4\linewidth,
at={(0in,0in)},
scale only axis,
xmin=0,
xmax=100,
xlabel style={font=\color{mycolor2}},
xlabel={number of outer iterations},
ymin=36.7,
ymax=37.05,
ylabel style={font=\color{mycolor2}},
ylabel={$\SNRfull$ [dB]},
axis background/.style={fill=white},
xmajorgrids,
ymajorgrids
]
\addplot [color=mycolor1, line width=1.0pt, mark=x, mark size=4pt, mark options={solid, mycolor1}, forget plot]
  table[row sep=crcr]{%
1	36.7349360747441\\
2	36.9620540533877\\
3	37.0426971267429\\
5	37.0452505505986\\
10	37.0377190334976\\
20	37.0279323297604\\
30	37.034052449662\\
50	37.0347691861738\\
75	37.0261278848646\\
100	37.0324545640969\\
};
\end{axis}
\end{tikzpicture}%

%% file: tex/outerIters-ODG.tex
%
%
\definecolor{mycolor1}{rgb}{0.06600,0.44300,0.74500}%
\definecolor{mycolor2}{rgb}{0.12941,0.12941,0.12941}%
\begin{tikzpicture}

\begin{axis}[%
width=0.45\linewidth,
height=0.4\linewidth,
at={(0in,0in)},
scale only axis,
xmin=0,
xmax=100,
xlabel style={font=\color{mycolor2}},
xlabel={number of outer iterations},
ymin=-1.14,
ymax=-0.94,
ylabel style={font=\color{mycolor2}},
ylabel={ODG},
axis background/.style={fill=white},
xmajorgrids,
ymajorgrids
]
\addplot [color=mycolor1, line width=1.0pt, mark=x, mark size=4pt, mark options={solid, mycolor1}, forget plot]
  table[row sep=crcr]{%
1	-1.12615209952304\\
2	-1.01680450802237\\
3	-0.984309344810705\\
5	-0.971129052147817\\
10	-0.960416343552834\\
20	-0.949626131672785\\
30	-0.959551382123605\\
50	-0.953541093453769\\
75	-0.949170638140075\\
100	-0.952786342338957\\
};
\end{axis}
\end{tikzpicture}%

%% file: tex/lambda_gcpa.tex
%
%
\definecolor{mycolor1}{rgb}{0.00000,0.44700,0.74100}%
\definecolor{mycolor3}{rgb}{0.00000,0.22350,0.5}
\begin{tikzpicture}

\begin{axis}[%
width=1\linewidth,
height=0.7\linewidth,
at={(0in,0in)},
scale only axis,
xmode=log,
xmin=1e-07,
xmax=100,
xlabel style={font=\color{white!15!black}},
xlabel={lambdas},
ymin=-1.8,
ymax=-0.9,
ylabel style={font=\color{white!15!black}},
ylabel={ODG},
axis background/.style={fill=white},
xmajorgrids,
ymajorgrids,
legend style={at={(0.03,0.97)}, anchor=north west, legend cell align=left, align=left, draw=white!15!black}
]
\addplot [color=mycolor1, line width=1.0pt, mark=x, mark size=4pt, mark options={solid, mycolor1}]
  table[row sep=crcr]{%
1e-07	-1.60601231309408\\
1e-06	-1.60561513852348\\
1e-05	-1.58540090191659\\
0.0001	-1.48530047188308\\
0.001	-1.24748910546481\\
0.01	-0.964798025250955\\
0.1	-1.04336662788788\\
1	-1.17246603962212\\
10	-1.39219267102956\\
100	-1.60704827670719\\
};
\addlegendentry{DPAI dataset}

\addplot [color=mycolor1, line width=1.0pt, dashed,draw opacity=0.75, mark=x, mark size=4pt, mark options={solid, mycolor1}]
  table[row sep=crcr]{%
1e-07	-1.72480203258738\\
1e-06	-1.72453444938823\\
1e-05	-1.71080992544387\\
0.0001	-1.66309963525207\\
0.001	-1.3912977066048\\
0.01	-1.00928961827477\\
0.1	-1.11478203488869\\
1	-1.29183525284385\\
10	-1.51346964110437\\
100	-1.66331883323381\\
};
\addlegendentry{IRMAS dataset}

\end{axis}

\end{tikzpicture}%

%% file: tex/snr_in_gaps_dpai.tex
%
%
\definecolor{mycolor5}{rgb}{0.12941,0.12941,0.12941}%

\definecolor{mycolor1}{rgb}{0.00000,0.44700,0.74100}%
\definecolor{mycolor2}{rgb}{0.85000,0.32500,0.09800}%
\definecolor{mycolor3}{rgb}{0.92900,0.69400,0.12500}%
\definecolor{mycolor4}{rgb}{0.49400,0.18400,0.55600}%
\begin{tikzpicture}

\begin{axis}[%
width=0.45\linewidth,
height=0.6\linewidth,
at={(0in,0in)},
scale only axis,
xmin=1,
xmax=6,
xtick={1, 2, 3, 4, 5, 6},
xlabel style={font=\color{mycolor5}},
xlabel={gap length},
ymin=0,
ymax=100,
ylabel style={font=\color{mycolor5}},
ylabel={$\TFSNR$ [dB]},
axis background/.style={fill=white},
xmajorgrids,
ymajorgrids,
legend style={legend cell align=left, align=left, draw=white!15!black, font=\footnotesize}
]

\addplot[area legend, draw=black, fill=mycolor1, draw opacity=0.1, fill opacity=0.2, forget plot]
table[row sep=crcr] {%
x	y\\
1	113.800737506944\\
2	77.5960811242252\\
3	16.025669851732\\
4	5.39239688746866\\
5	3.25686766466567\\
6	1.79140416631555\\
6	6.66157987506339\\
5	9.10971417392859\\
4	11.2363040529617\\
3	23.368948239062\\
2	86.9688972773413\\
1	120.333083451484\\
}--cycle;
\addplot [color=mycolor1, line width=1.0pt]
  table[row sep=crcr]{%
1	117.066910479214\\
2	82.2824892007832\\
3	19.697309045397\\
4	8.31435047021518\\
5	6.18329091929713\\
6	4.22649202068947\\
};
\addlegendentry{U-PHAIN-TF}

\addplot[area legend, draw=black, fill=mycolor2, draw opacity=0.1, fill opacity=0.2, forget plot]
table[row sep=crcr] {%
x	y\\
1	126.592123592144\\
2	81.0336503885144\\
3	15.2901556917994\\
4	5.68698435390809\\
5	2.81054884086684\\
6	2.30133564860382\\
6	3.38260012023793\\
5	5.64324957385769\\
4	7.08098384667327\\
3	21.9037005155653\\
2	87.7514826589203\\
1	132.207197415551\\
}--cycle;
\addplot [color=mycolor2, dashdotted, line width=1.0pt]
  table[row sep=crcr]{%
1	129.399660503848\\
2	84.3925665237174\\
3	18.5969281036824\\
4	6.38398410029068\\
5	4.22689920736227\\
6	2.84196788442088\\
};
\addlegendentry{{$\ell_1$}-TF}

\addplot[area legend, draw=black, fill=mycolor3, draw opacity=0.1, fill opacity=0.2, forget plot]
table[row sep=crcr] {%
x	y\\
1	10.5284884343165\\
2	7.17869387798298\\
3	3.47915736747889\\
4	1.21043964213257\\
5	1.13897589376065\\
6	0.57577241719681\\
6	3.56683740075827\\
5	4.06010674410329\\
4	4.31493566325689\\
3	8.19118098427644\\
2	15.5264766121779\\
1	19.782570035227\\
}--cycle;
\addplot [color=mycolor3, dotted, line width=1.0pt]
  table[row sep=crcr]{%
1	15.1555292347718\\
2	11.3525852450804\\
3	5.83516917587766\\
4	2.76268765269473\\
5	2.59954131893197\\
6	2.07130490897754\\
};
\addlegendentry{DPAI}

\addplot[area legend, draw=black, fill=mycolor4, draw opacity=0.1, fill opacity=0.2, forget plot]
table[row sep=crcr] {%
x	y\\
1	47.0862275993188\\
2	36.6868741794924\\
3	12.1925404694833\\
4	4.43221433087778\\
5	3.26166065674412\\
6	1.91147289204921\\
6	7.41791932826385\\
5	10.068699775222\\
4	12.7104871836808\\
3	23.7420802361299\\
2	54.569765690261\\
1	66.3299994772296\\
}--cycle;
\addplot [color=mycolor4, dashed, line width=1.0pt]
  table[row sep=crcr]{%
1	56.7081135382742\\
2	45.6283199348767\\
3	17.9673103528066\\
4	8.57135075727929\\
5	6.66518021598308\\
6	4.66469611015653\\
};
\addlegendentry{Janssen-TF}

\end{axis}
\end{tikzpicture}%

%% file: tex/odg_dpai_withL1.tex
%
%
\definecolor{mycolor1}{rgb}{0.00000,0.44700,0.74100}%
\definecolor{mycolor2}{rgb}{0.85000,0.32500,0.09800}%
\definecolor{mycolor3}{rgb}{0.92900,0.69400,0.12500}%
\definecolor{mycolor4}{rgb}{0.49400,0.18400,0.55600}%
\definecolor{mycolor5}{rgb}{0.12941,0.12941,0.12941}%
\begin{tikzpicture}

\begin{axis}[%
width=0.45\linewidth,
height=0.6\linewidth,
at={(0in,0in)},
scale only axis,
xmin=1,
xmax=6,
xtick={1, 2, 3, 4, 5, 6},
xlabel style={font=\color{mycolor5}},
xlabel={gap length},
ymin=-3.5,
ymax=0,
ylabel style={font=\color{mycolor5}},
ylabel={ODG},
axis background/.style={fill=white},
xmajorgrids,
ymajorgrids,
]

\addplot[area legend, draw=black, fill=mycolor1, draw opacity=0.1, fill opacity=0.2, forget plot]
table[row sep=crcr] {%
x	y\\
1	-1.2770981887067e-08\\
2	-6.65666385518477e-05\\
3	-0.125330669988479\\
4	-1.52018975239912\\
5	-2.60502515379664\\
6	-2.84244432301318\\
6	-2.06123924969387\\
5	-1.710037397279\\
4	-0.59476151974671\\
3	-0.0323415760412622\\
2	1.37726611984873e-05\\
1	4.94565216951911e-09\\
}--cycle;
\addplot [color=mycolor1, line width=1.0pt]
  table[row sep=crcr]{%
1	-3.91266485877395e-09\\
2	-2.63969886766802e-05\\
3	-0.0788361230148706\\
4	-1.05747563607291\\
5	-2.15753127553782\\
6	-2.45184178635353\\
};

\addplot[area legend, draw=black, fill=mycolor2, draw opacity=0.1, fill opacity=0.2, forget plot]
table[row sep=crcr] {%
x	y\\
1	-7.27946547511848e-11\\
2	-1.20152655836235e-05\\
3	-0.142439017857189\\
4	-2.15418570936715\\
5	-3.18490369339112\\
6	-3.40601650534862\\
6	-3.13457689086657\\
5	-2.93815603701555\\
4	-1.09110474724212\\
3	-0.0453978966147007\\
2	2.94439474161395e-06\\
1	-7.87326203965942e-12\\
}--cycle;
\addplot [color=mycolor2, dashdotted, line width=1.0pt]
  table[row sep=crcr]{%
1	-4.03339583954221e-11\\
2	-4.53543542100476e-06\\
3	-0.0939184572359446\\
4	-1.62264522830464\\
5	-3.06152986520333\\
6	-3.2702966981076\\
};

\addplot[area legend, draw=black, fill=mycolor3, draw opacity=0.1, fill opacity=0.2, forget plot]
table[row sep=crcr] {%
x	y\\
1	-0.73003605681747\\
2	-1.60159560422825\\
3	-2.38636510394289\\
4	-2.9834870047098\\
5	-3.16873044363996\\
6	-3.0796235656356\\
6	-2.80967887379931\\
5	-2.74013321320186\\
4	-2.45861267729442\\
3	-1.22155651177496\\
2	-0.277379153901612\\
1	-0.290994063286441\\
}--cycle;
\addplot [color=mycolor3, dotted, line width=1.0pt]
  table[row sep=crcr]{%
1	-0.510515060051956\\
2	-0.939487379064929\\
3	-1.80396080785893\\
4	-2.72104984100211\\
5	-2.95443182842091\\
6	-2.94465121971745\\
};

\addplot[area legend, draw=black, fill=mycolor4, draw opacity=0.1, fill opacity=0.2, forget plot]
table[row sep=crcr] {%
x	y\\
1	-0.00617339519671911\\
2	-0.0830415109824161\\
3	-1.06574917671802\\
4	-2.59492534525323\\
5	-3.09929306583604\\
6	-3.15459739122956\\
6	-2.74558657216464\\
5	-2.23223042634096\\
4	-1.15133445523747\\
3	-0.11337828971999\\
2	0.0171102299842962\\
1	-0.000557472835509331\\
}--cycle;
\addplot [color=mycolor4, dashed, line width=1.0pt]
  table[row sep=crcr]{%
1	-0.00336543401611422\\
2	-0.0329656404990599\\
3	-0.589563733219007\\
4	-1.87312990024535\\
5	-2.6657617460885\\
6	-2.9500919816971\\
};

\end{axis}

\end{tikzpicture}%

%% file: tex/snr_in_gaps_irmas.tex
%
%
\definecolor{mycolor1}{rgb}{0.00000,0.44700,0.74100}%
\definecolor{mycolor2}{rgb}{0.85000,0.32500,0.09800}%
\definecolor{mycolor3}{rgb}{0.92900,0.69400,0.12500}%
\definecolor{mycolor4}{rgb}{0.49400,0.18400,0.55600}%
\definecolor{mycolor5}{rgb}{0.12941,0.12941,0.12941}%
\begin{tikzpicture}

\begin{axis}[%
width=0.45\linewidth,
height=0.6\linewidth,
at={(0in,0in)},
scale only axis,
xmin=1,
xmax=6,
xtick={1, 2, 3, 4, 5, 6},
xlabel style={font=\color{mycolor5}},
xlabel={gap length},
ymin=0,
ymax=100,
ylabel style={font=\color{mycolor5}},
ylabel={ $\TFSNR$ [dB]},
axis background/.style={fill=white},
xmajorgrids,
ymajorgrids,
legend style={legend cell align=left, align=left, draw=white!15!black, font=\footnotesize}
]

\addplot[area legend, draw=black, fill=mycolor1, draw opacity=0.1, fill opacity=0.2, forget plot]
table[row sep=crcr] {%
x	y\\
1	135.925812844012\\
2	77.8909258756127\\
3	21.9192394923692\\
4	10.3565389316704\\
5	6.36113093661259\\
6	4.72387815415711\\
6	5.68032677072726\\
5	7.90300050427335\\
4	12.4547526545194\\
3	24.0891353351201\\
2	78.9510225118388\\
1	143.895866891537\\
}--cycle;
\addplot [color=mycolor1, line width=1.0pt]
  table[row sep=crcr]{%
1	139.910839867774\\
2	78.4209741937258\\
3	23.0041874137447\\
4	11.4056457930949\\
5	7.13206572044297\\
6	5.20210246244219\\
};
\addlegendentry{U-PHAIN-TF}

\addplot[area legend, draw=black, fill=mycolor2, draw opacity=0.1, fill opacity=0.2, forget plot]
table[row sep=crcr] {%
x	y\\
1	193.85643133812\\
2	78.4022298564508\\
3	19.0665806537016\\
4	6.99181270512238\\
5	4.33635712366941\\
6	3.36116535107845\\
6	3.67758513622214\\
5	4.66450911145308\\
4	7.4784674690868\\
3	20.6750693283512\\
2	79.3147665931989\\
1	224.486979728785\\
}--cycle;
\addplot [color=mycolor2, dashdotted, line width=1.0pt]
  table[row sep=crcr]{%
1	209.171705533453\\
2	78.8584982248249\\
3	19.8708249910264\\
4	7.23514008710459\\
5	4.50043311756125\\
6	3.5193752436503\\
};
\addlegendentry{$\ell_1$-TF}

\addplot[area legend, draw=black, fill=mycolor3, draw opacity=0.1, fill opacity=0.2, forget plot]
table[row sep=crcr] {%
x	y\\
1	17.5030930629514\\
2	9.46758185561025\\
3	5.277653357422\\
4	4.1761770634435\\
5	2.84439287887391\\
6	2.41225147740066\\
6	3.33480285266121\\
5	3.84998241306245\\
4	5.58280198134035\\
3	6.64641795106408\\
2	11.206025013665\\
1	19.7836883754044\\
}--cycle;
\addplot [color=mycolor3, dotted, line width=1.0pt]
  table[row sep=crcr]{%
1	18.6433907191779\\
2	10.3368034346376\\
3	5.96203565424304\\
4	4.87948952239192\\
5	3.34718764596818\\
6	2.87352716503094\\
};
\addlegendentry{DPAI}

\addplot[area legend, draw=black, fill=mycolor4, draw opacity=0.1, fill opacity=0.2, forget plot]
table[row sep=crcr] {%
x	y\\
1	49.9799564226058\\
2	39.987422083407\\
3	20.2246475821885\\
4	10.8750070016745\\
5	6.71978011354308\\
6	5.07141037482561\\
6	6.01825704822641\\
5	8.24554548736508\\
4	13.0123010547528\\
3	23.3401060285503\\
2	44.7209717511774\\
1	54.356819151225\\
}--cycle;
\addplot [color=mycolor4, dashed, line width=1.0pt]
  table[row sep=crcr]{%
1	52.1683877869154\\
2	42.3541969172922\\
3	21.7823768053694\\
4	11.9436540282137\\
5	7.48266280045408\\
6	5.54483371152601\\
};
\addlegendentry{Janssen-TF}

\end{axis}

\end{tikzpicture}%

%% file: tex/odg_irmas_withL1.tex
%
%
\definecolor{mycolor1}{rgb}{0.00000,0.44700,0.74100}%
\definecolor{mycolor2}{rgb}{0.85000,0.32500,0.09800}%
\definecolor{mycolor3}{rgb}{0.92900,0.69400,0.12500}%
\definecolor{mycolor4}{rgb}{0.49400,0.18400,0.55600}%
\definecolor{mycolor5}{rgb}{0.12941,0.12941,0.12941}%
\begin{tikzpicture}

\begin{axis}[%
width=0.45\linewidth,
height=0.6\linewidth,
at={(0in,0in)},
scale only axis,
xmin=1,
xmax=6,
xtick={1, 2, 3, 4, 5, 6},
xlabel style={font=\color{mycolor5}},
xlabel={gap length},
ymin=-3.5,
ymax=0,
ylabel style={font=\color{mycolor5}},
ylabel={ODG},
axis background/.style={fill=white},
xmajorgrids,
ymajorgrids,
]

\addplot[area legend, draw=black, fill=mycolor1, draw opacity=0.1, fill opacity=0.2, forget plot]
table[row sep=crcr] {%
x	y\\
1	-4.66187192587039e-09\\
2	-0.000307294382909369\\
3	-0.0869718126283407\\
4	-1.13108043334902\\
5	-2.38206871763511\\
6	-2.71201582197956\\
6	-2.57404122229386\\
5	-2.18325563830372\\
4	-0.972508630915937\\
3	-0.0691142491941976\\
2	-5.55776999261202e-05\\
1	1.02573209525866e-10\\
}--cycle;
\addplot [color=mycolor1, line width=1.0pt]
  table[row sep=crcr]{%
1	-2.27964935817226e-09\\
2	-0.000181436041417745\\
3	-0.0780430309112692\\
4	-1.05179453213248\\
5	-2.28266217796941\\
6	-2.64302852213671\\
};

\addplot[area legend, draw=black, fill=mycolor2, draw opacity=0.1, fill opacity=0.2, forget plot]
table[row sep=crcr] {%
x	y\\
1	-2.16658701152248e-11\\
2	-9.64579096511768e-05\\
3	-0.223913252245013\\
4	-1.96994410684215\\
5	-3.01606370899627\\
6	-3.3439049131987\\
6	-3.32031063810858\\
5	-2.95723962253154\\
4	-1.77378951382916\\
3	-0.11764533248104\\
2	-1.1049998104593e-05\\
1	-4.1315681444386e-12\\
}--cycle;
\addplot [color=mycolor2, dashdotted, line width=1.0pt]
  table[row sep=crcr]{%
1	-1.28987191298317e-11\\
2	-5.37539538778849e-05\\
3	-0.170779292363027\\
4	-1.87186681033566\\
5	-2.98665166576391\\
6	-3.33210777565364\\
};
\addplot[area legend, draw=black, fill=mycolor3, draw opacity=0.1, fill opacity=0.2, forget plot]
table[row sep=crcr] {%
x	y\\
1	-0.600062700481302\\
2	-1.38157820234933\\
3	-2.26837475327441\\
4	-2.62938647362694\\
5	-2.92108560827893\\
6	-3.01078152911988\\
6	-2.93952552797178\\
5	-2.83294946844038\\
4	-2.45274452274243\\
3	-1.99482443859493\\
2	-1.00053639237388\\
1	-0.356001389016841\\
}--cycle;
\addplot [color=mycolor3, dotted, line width=1.0pt]
  table[row sep=crcr]{%
1	-0.478032044749071\\
2	-1.1910572973616\\
3	-2.13159959593467\\
4	-2.54106549818468\\
5	-2.87701753835966\\
6	-2.97515352854583\\
};

\addplot[area legend, draw=black, fill=mycolor4, draw opacity=0.1, fill opacity=0.2, forget plot]
table[row sep=crcr] {%
x	y\\
1	-0.0562758441790677\\
2	-0.290440265780606\\
3	-1.52882823617355\\
4	-2.63804012357358\\
5	-3.01249038455502\\
6	-3.14534720676114\\
6	-3.08367196635815\\
5	-2.91272242673117\\
4	-2.44116408118509\\
3	-1.26094987774165\\
2	-0.16605106176648\\
1	-0.0244502428316337\\
}--cycle;
\addplot [color=mycolor4, dashed, line width=1.0pt]
  table[row sep=crcr]{%
1	-0.0403630435053507\\
2	-0.228245663773543\\
3	-1.3948890569576\\
4	-2.53960210237934\\
5	-2.96260640564309\\
6	-3.11450958655964\\
};

\end{axis}

\end{tikzpicture}%

%% file: tex/snrScatter1.tex
%
%
\definecolor{mycolor1}{rgb}{0.06600,0.44300,0.74500}%
\definecolor{mycolor2}{rgb}{0.86600,0.32900,0.00000}%
\definecolor{mycolor3}{rgb}{0.12941,0.12941,0.12941}%
\begin{tikzpicture}

\begin{axis}[%
width=0.45\linewidth,
height=0.4\linewidth,
at={(0in,0in)},
scale only axis,
xmin=-30.0466763345926,
xmax=134.000963731495,
xlabel style={font=\color{mycolor3}},
xlabel={U-PHAIN-TF SNR [dB]},
ymin=0,
ymax=103.954287396902,
ylabel style={font=\color{mycolor3}},
ylabel={Janssen-TF $\SNRfull$ [dB]},
axis background/.style={fill=white},
title style={font=\bfseries\color{mycolor3}, at={(0.5in,1in)},},
title={Mask 1},
axis x line*=bottom,
axis y line*=left,
xmajorgrids,
ymajorgrids
]
\addplot[only marks, mark=x, mark options={solid}, mark size=3pt, draw=mycolor1, forget plot] table[row sep=crcr]{%
x	y\\
93.5486999858741	56.7419225150042\\
98.3616032503724	56.0479559301518\\
90.2608998201992	60.9757379031133\\
94.2314017117072	56.9856152781137\\
92.0684182824943	63.1260707007537\\
90.3549853562138	80.026301292193\\
94.8267174461877	47.6031999060971\\
94.3797091395016	56.7201915361593\\
93.4445381537214	50.9796704228228\\
92.9736346814446	59.5614061061687\\
98.1053667514659	55.4200804214029\\
93.6309373260224	57.3392410178389\\
88.6293010424287	73.6929495220163\\
89.8099242428085	72.419945135336\\
87.8981166154933	73.8630330587923\\
91.8180934468047	59.1051524367772\\
94.1710330623441	73.0084000257565\\
91.5998088077482	63.6576455396806\\
92.5250460359631	72.3631159319308\\
92.8269844949861	56.8101749833434\\
92.8341959790787	53.6124759411076\\
91.087344886425	70.4150348301862\\
98.7661452162991	57.3163406349655\\
92.2722369267663	74.5177332850552\\
88.6111521551584	68.2940441840748\\
93.7861723449043	63.9413800634258\\
94.9111385668986	60.6579276651745\\
89.4302372002324	57.0833460735701\\
90.7453064640675	69.6189424703745\\
92.5207718479499	58.0527563941261\\
89.651535320497	69.6950876127589\\
91.3362858715502	61.6761671865091\\
91.0334227572598	75.595307108879\\
92.754613678561	63.0468730579541\\
94.3678192166965	56.9948140572171\\
103.954287396902	60.8055959286476\\
89.8237309843271	82.2394060576271\\
92.9684193273095	63.5595449728729\\
99.9794693669257	58.3007788449184\\
99.6335290662718	56.6311275265662\\
99.8624052422454	54.9183123455285\\
97.0288766074159	58.527035780371\\
92.3769881755276	56.6141375238046\\
86.9724591797379	59.8688652796384\\
91.6514373454168	77.2942506038265\\
89.9727760761045	68.0396219313042\\
91.4687282514183	65.7332475032313\\
92.0489935773666	62.4098386181015\\
95.0644087188447	59.0394125907665\\
93.7822305765522	56.2920051354012\\
90.6415931831796	61.5040520862917\\
91.8451580229359	54.1532202457572\\
92.8579330445191	57.4154853738848\\
91.7790095997517	60.0002060627621\\
86.6740636818734	60.1757290126586\\
98.1402094617171	53.4467327691161\\
91.6472141475476	67.088218092404\\
91.1723490505635	84.9741974526649\\
93.2595677778754	75.4505887586323\\
92.3762086375588	62.9825431174617\\
};
\addplot [color=mycolor3,draw opacity=0.5,line width=1.0pt, forget plot]
  table[row sep=crcr]{%
0	0\\
103.954287396902	103.954287396902\\
};
\end{axis}
\end{tikzpicture}%

%% file: tex/snrScatter2.tex
%
%
\definecolor{mycolor1}{rgb}{0.06600,0.44300,0.74500}%
\definecolor{mycolor2}{rgb}{0.86600,0.32900,0.00000}%
\definecolor{mycolor3}{rgb}{0.12941,0.12941,0.12941}%
\begin{tikzpicture}

\begin{axis}[%
width=0.45\linewidth,
height=0.4\linewidth,
at={(0in,0in)},
scale only axis,
xmin=-27.0919519470751,
xmax=120.823602246343,
xlabel style={font=\color{mycolor3}},
xlabel={U-PHAIN-TF SNR [dB]},
ymin=0,
ymax=93.7316502992679,
ylabel style={font=\color{mycolor3}},
axis background/.style={fill=white},
title style={font=\bfseries\color{mycolor3}, at={(0.5in,1in)},},
title={Mask 2},
axis x line*=bottom,
axis y line*=left,
xmajorgrids,
ymajorgrids
]
\addplot[only marks, mark=x, mark options={solid}, mark size=3pt, draw=mycolor1, forget plot] table[row sep=crcr]{%
x	y\\
90.3505842043644	52.3142808400794\\
92.6140488362326	42.9413167254915\\
87.1731065278677	55.7072215267912\\
90.078699711843	40.5163754240296\\
89.4649738016928	58.4310635617799\\
87.227475344522	74.1243788881496\\
88.7515572070425	38.9570928206696\\
91.6771326063433	42.8981023441708\\
87.5784684621009	38.4272848717548\\
90.7098245524074	54.3154368978813\\
89.4450809193503	41.350227776324\\
91.2228512660402	42.5908380301452\\
86.2825219204451	63.9300560938562\\
87.7203769647404	59.0761404068631\\
85.6065422389302	52.6778638978147\\
87.8215736844454	56.0790189336731\\
90.0698901357108	54.67395794942\\
88.7196520130584	52.7133095796228\\
90.5491800733308	63.6547874023372\\
90.4379558541631	51.6383322430307\\
89.4796878832594	40.9357813362637\\
88.8739602603678	62.51811328547\\
92.2502554051878	43.154849821894\\
90.2517984811874	52.1779650058639\\
86.4750862268382	61.3457916261619\\
90.6019396395061	54.8117570680539\\
90.4482859470966	46.3893318045239\\
86.4554180811787	49.9160873444868\\
88.679791134583	53.9624924572259\\
89.1433744595183	50.4825803158415\\
86.7979116864112	63.1334358512551\\
88.8319043002943	51.0832624232655\\
88.8146766062128	67.3669216043748\\
90.5165539403421	55.9882311240483\\
88.4617158511074	48.4214786046938\\
91.0988346416724	45.7078723726456\\
87.3920453299655	75.966745468643\\
90.368741089049	60.8740409437665\\
91.1127596894356	44.6970551557677\\
93.7316502992679	44.7955798697173\\
92.1484360099032	44.8222418751567\\
91.3570149129581	45.6243063470091\\
89.8599343297081	47.8339107843901\\
85.0558477535233	53.3958707277669\\
89.0797887576066	69.5220493197614\\
87.9770525270524	59.6690568227242\\
88.9517876046673	58.0216241643603\\
88.8315008463657	52.1665551356917\\
91.5073928415518	50.6424178094842\\
89.8998696732783	45.6517308622701\\
88.0905222722614	60.5766473547998\\
88.2866293099088	42.9772469360395\\
89.9481423208237	53.5483076691343\\
89.1751181018845	51.0651699210301\\
84.5667123293387	62.3264611807016\\
91.4777375622978	44.5748994064115\\
89.3188663065732	61.5869908786058\\
88.0867618421972	75.6432280696727\\
91.1052192028416	66.2167135709101\\
89.8891574262665	57.7706009358695\\
};
\addplot [color=mycolor3,draw opacity=0.5,line width=1.0pt, forget plot]
  table[row sep=crcr]{%
0	0\\
93.7316502992679	93.7316502992679\\
};
\end{axis}
\end{tikzpicture}%

%% file: tex/snrScatter3.tex
%
%
\definecolor{mycolor1}{rgb}{0.06600,0.44300,0.74500}%
\definecolor{mycolor2}{rgb}{0.86600,0.32900,0.00000}%
\definecolor{mycolor3}{rgb}{0.12941,0.12941,0.12941}%
\begin{tikzpicture}

\begin{axis}[%
width=0.45\linewidth,
height=0.4\linewidth,
at={(0in,0in)},
scale only axis,
xmin=-12.7246048122827,
xmax=56.748682914563,
xlabel style={font=\color{mycolor3}},
xlabel={U-PHAIN-TF SNR [dB]},
ymin=0,
ymax=44.0240781022804,
ylabel style={font=\color{mycolor3}},
axis background/.style={fill=white},
title style={font=\bfseries\color{mycolor3}, at={(0.5in,1in)},},
title={Mask 3},
axis x line*=bottom,
axis y line*=left,
xmajorgrids,
ymajorgrids
]
\addplot[only marks, mark=x, mark options={solid}, mark size=3pt, draw=mycolor1, forget plot] table[row sep=crcr]{%
x	y\\
34.6072271901607	34.6307613379312\\
29.3305300394755	24.0652604133536\\
32.5779879649693	31.3758588766608\\
29.7499240588506	24.5365675428197\\
40.1056448900889	38.3438651740403\\
40.7052310559212	44.0240781022804\\
25.4976441142365	22.9073538108644\\
29.8479501923199	25.761449284261\\
26.598264585434	20.2260589107344\\
33.2206380705989	31.0426581631022\\
30.4255289781338	26.0153869418153\\
30.8642611765728	26.1635178325815\\
40.682622984829	41.809233059268\\
26.702391523542	25.8819917145951\\
28.5365947215816	28.7938749895085\\
39.6107541696353	37.053630317119\\
36.6534328274723	34.6575464034553\\
36.009640838748	34.8664425222221\\
35.0885794079235	40.9107175951861\\
34.1510417301246	32.2880810371665\\
29.8389189284166	24.3346928149788\\
34.1055606410747	35.7344032593461\\
29.5959662782469	23.9006971241475\\
30.7897505792905	35.8268235863668\\
35.4061267087563	34.962488047564\\
34.4428341213266	30.8068124775492\\
24.6089806303026	21.5199017168318\\
31.8144100807772	25.6182484774794\\
35.5743846059254	36.3687952036791\\
33.2036449240343	30.7690423393891\\
42.704577847748	43.1061093023814\\
33.5322613775518	30.6862511849186\\
29.699270278629	32.6034378697868\\
32.6699330821927	34.6163494692116\\
33.4542299781817	31.3142419036997\\
32.0933915668879	26.9522060256876\\
37.0981603165417	39.5157838109067\\
32.110295937702	33.7248932354172\\
30.1904076016122	24.5003127963971\\
29.7673107957922	25.6839727075139\\
24.5684493849194	24.3273095682749\\
33.3825038748304	32.0265098663987\\
34.1393146852217	31.1616353647704\\
23.6276378577666	26.50737412724\\
39.9930972714567	41.046584146393\\
34.5911279390497	36.5212624372372\\
35.7376333309477	35.5257687246893\\
35.1126142967714	33.4163246123096\\
28.5718303143493	26.4975549425692\\
29.4365202231939	23.569336352906\\
32.5958748194764	31.998926304814\\
31.7397862826515	28.7640716899068\\
35.700779201492	34.2528449919242\\
31.9083018265852	31.7668339631771\\
35.5988400757988	36.5366111586735\\
27.6118169078351	24.0311654554054\\
38.8772551278187	41.4614852284071\\
36.7459876946956	37.0276151429552\\
38.398056445797	40.1412156353562\\
35.5651766203711	36.4502820668924\\
};
\addplot [color=mycolor3,draw opacity=0.5,line width=1.0pt, forget plot]
  table[row sep=crcr]{%
0	0\\
44.0240781022804	44.0240781022804\\
};
\end{axis}
\end{tikzpicture}%

%% file: tex/snrScatter4.tex
%
%
\definecolor{mycolor1}{rgb}{0.06600,0.44300,0.74500}%
\definecolor{mycolor2}{rgb}{0.86600,0.32900,0.00000}%
\definecolor{mycolor3}{rgb}{0.12941,0.12941,0.12941}%
\begin{tikzpicture}

\begin{axis}[%
width=0.45\linewidth,
height=0.4\linewidth,
at={(0in,0in)},
scale only axis,
xmin=-9.36639920156577,
xmax=41.7718920298257,
xlabel style={font=\color{mycolor3}},
xlabel={U-PHAIN-TF SNR [dB]},
ymin=0,
ymax=32.4054928282599,
ylabel style={font=\color{mycolor3}},
ylabel={Janssen-TF $\SNRfull$ [dB]},
axis background/.style={fill=white},
title style={font=\bfseries\color{mycolor3}, at={(0.5in,1in)},},
title={Mask 4},
axis x line*=bottom,
axis y line*=left,
xmajorgrids,
ymajorgrids
]
\addplot[only marks, mark=x, mark options={solid}, mark size=3pt, draw=mycolor1, forget plot] table[row sep=crcr]{%
x	y\\
21.9956320051076	21.7177918943539\\
16.6774574998667	17.8139014776984\\
26.6188545550785	27.0427296747173\\
18.1369618645599	16.875609770192\\
20.4058016168729	25.5701895154271\\
22.4566037676407	20.4025701966717\\
14.8190124678315	13.159109895428\\
17.1722696523833	16.0444987151332\\
18.2460488406883	17.9790203096236\\
20.3312357474223	19.7137401745209\\
19.5854021909626	19.0382405638303\\
18.7696428850898	17.9263912569804\\
15.9475339892797	20.3966388063871\\
27.6416199812463	26.9389413826462\\
24.4298625768556	24.0924930940928\\
32.4054928282599	31.4104508276857\\
15.0051349361022	20.0740728983658\\
24.3716537124483	26.7781316706858\\
21.2123224129306	19.6279752894469\\
20.2130060837694	21.8951705473809\\
19.7252030056549	18.5019458314882\\
22.8647313168934	22.0884061555347\\
18.1156499950317	18.0406140354411\\
27.5108549687197	28.1728979950303\\
20.1808962261116	24.7668785032484\\
22.8448154627158	24.333461356196\\
16.7050369269902	15.6498135779437\\
15.7220447681408	15.6582312851411\\
17.0869547285101	20.5880338776545\\
21.1194601466073	21.8101130064632\\
31.711844321948	30.5841049190303\\
17.8884143424268	18.1311120632233\\
22.2597124774715	25.4654658795602\\
18.2648213515845	18.9008797131079\\
18.6051216048175	19.5111510008084\\
17.4856173842869	17.2875888979278\\
26.6131157008315	20.4365399412537\\
17.8933167131729	20.2517220653275\\
18.2530185971156	17.5297737470201\\
18.2453598367445	18.622492730779\\
14.7941505526372	15.0675618902141\\
24.9350665022915	25.3792658415912\\
20.1102033064406	20.4875634081723\\
15.023295762	14.4360812149207\\
20.1969793016693	22.3360704371249\\
18.7856174996197	22.991835239208\\
24.1913970160542	26.9105786289352\\
21.3112599692966	20.7282518306701\\
16.7961363395622	16.9301170268794\\
19.4306500245465	20.0175623016564\\
18.2123189082605	22.1875738244769\\
20.1648475584427	20.7153135031175\\
21.8826279659799	24.7184357789578\\
17.911597694648	19.120155287999\\
25.6508070210062	26.4857404768749\\
18.0271218460872	18.7624881163772\\
17.9179922515336	21.3529672111116\\
18.8580557027644	17.1780601615075\\
20.93591986687	21.9591968671813\\
21.7776685278121	21.7798871123923\\
};
\addplot [color=mycolor3,draw opacity=0.5,line width=1.0pt, forget plot]
  table[row sep=crcr]{%
0	0\\
32.4054928282599	32.4054928282599\\
};
\end{axis}
\end{tikzpicture}%

%% file: tex/snrScatter5.tex
%
%
\definecolor{mycolor1}{rgb}{0.06600,0.44300,0.74500}%
\definecolor{mycolor2}{rgb}{0.86600,0.32900,0.00000}%
\definecolor{mycolor3}{rgb}{0.12941,0.12941,0.12941}%
\begin{tikzpicture}

\begin{axis}[%
width=0.45\linewidth,
height=0.4\linewidth,
at={(0in,0in)},
scale only axis,
xmin=-7.55057545567951,
xmax=33.6737540126382,
xlabel style={font=\color{mycolor3}},
xlabel={U-PHAIN-TF SNR [dB]},
ymin=0,
ymax=26.1231785569587,
ylabel style={font=\color{mycolor3}},
axis background/.style={fill=white},
title style={font=\bfseries\color{mycolor3}, at={(0.5in,1in)},},
title={Mask 5},
axis x line*=bottom,
axis y line*=left,
xmajorgrids,
ymajorgrids
]
\addplot[only marks, mark=x, mark options={solid}, mark size=3pt, draw=mycolor1, forget plot] table[row sep=crcr]{%
x	y\\
17.0662717201258	17.454737832579\\
13.7280908873547	14.3248501838049\\
15.6859352130489	16.7525170145387\\
12.0784845165819	11.4715946996727\\
16.9082731127353	20.4089988017978\\
16.0947139556011	13.5116815994034\\
12.1545908881786	12.6324849383176\\
12.0227371752693	12.9200990985521\\
14.2650376123925	13.4892962128288\\
12.8831655361934	13.1292462385783\\
14.1399900385498	13.0902360924466\\
11.967391230454	10.8374189845596\\
23.6267198609388	20.874147698435\\
14.4963424866037	15.9633441070166\\
16.4075363362222	17.0441673441623\\
19.7963887081864	20.0964700818186\\
11.8180138181399	16.3488507101176\\
20.0948657380562	18.6913565160544\\
16.0273132289728	14.0519981115134\\
13.1825977777869	16.5581645611908\\
11.5770675674472	11.3705619243075\\
23.0380705265898	17.3108908599403\\
12.6812207079434	13.9629267839922\\
20.4588017027071	17.3739949451023\\
18.4727045231095	18.3786325681134\\
13.9801972455241	12.7421505658171\\
14.7163435837432	16.226971555307\\
13.3734922566129	14.2243926589553\\
11.8882091610256	14.1650494080602\\
13.5964475545696	15.0881629248525\\
24.6267982446607	26.1231785569587\\
11.8875717372461	12.7284411656677\\
13.7061022974155	15.0274538751218\\
13.8069422865722	15.0089110206855\\
11.739410143371	14.4008257547003\\
16.9761359874295	16.7104228684975\\
19.2230914820139	18.0338453227864\\
13.1457247404228	14.1332279200502\\
10.9279223521414	10.3691672122612\\
11.8445820949172	12.3146742363837\\
11.7838445684916	11.4962205171684\\
17.1683800760407	19.50169503\\
12.8739904927869	14.0487235126451\\
13.968806409175	14.0885971807127\\
18.302258978229	15.1553412864975\\
13.7924950723908	14.9857178492828\\
20.94298710982	23.4273369618696\\
14.9771743238607	14.7355374010002\\
12.642565417296	13.4182379859323\\
12.7724035055007	12.686390222328\\
13.3745661654064	14.7558505028523\\
13.9339230190098	12.2303300424791\\
16.1618243755056	19.0503813443168\\
14.7102034658869	15.5568504111519\\
16.1865581781549	18.1898431274597\\
12.5037952383117	12.2747165853586\\
13.2033651662139	17.262623855169\\
13.2296281385234	12.5683862052485\\
13.8879940709824	13.7392451433416\\
16.6807897146683	18.5312855762212\\
};
\addplot [color=mycolor3,draw opacity=0.5,line width=1.0pt, forget plot]
  table[row sep=crcr]{%
0	0\\
26.1231785569587	26.1231785569587\\
};
\end{axis}
\end{tikzpicture}%

%% file: tex/snrScatter6.tex
%
%
\definecolor{mycolor1}{rgb}{0.06600,0.44300,0.74500}%
\definecolor{mycolor2}{rgb}{0.86600,0.32900,0.00000}%
\definecolor{mycolor3}{rgb}{0.12941,0.12941,0.12941}%
\begin{tikzpicture}

\begin{axis}[%
width=0.45\linewidth,
height=0.4\linewidth,
at={(0in,0in)},
scale only axis,
xmin=-5.86374688401117,
xmax=26.1509035071017,
xlabel style={font=\color{mycolor3}},
xlabel={U-PHAIN-TF SNR [dB]},
ymin=0,
ymax=20.2871566230906,
ylabel style={font=\color{mycolor3}},
axis background/.style={fill=white},
title style={font=\bfseries\color{mycolor3}, at={(0.5in,1in)},},
title={Mask 6},
axis x line*=bottom,
axis y line*=left,
xmajorgrids,
ymajorgrids
]
\addplot[only marks, mark=x, mark options={solid}, mark size=3pt, draw=mycolor1, forget plot] table[row sep=crcr]{%
x	y\\
15.0497605864096	17.17852330763\\
11.5765736755705	12.6020435151933\\
14.532308517249	13.9940210972924\\
10.8492411043165	9.85695944947515\\
11.1870276634051	13.5450520379383\\
11.0363683808488	10.0489228384318\\
12.0705003533568	10.5079711297732\\
11.822065962519	12.2881263347033\\
9.04347089684212	8.53948975013969\\
9.83411320459508	10.5226872775654\\
14.4379609033764	15.6413089352579\\
10.4920621610503	10.4658117011556\\
12.9352133070339	13.893119459015\\
13.3296005925273	14.1983146072476\\
10.2288809194025	11.2336275697274\\
16.3860271358703	16.6171902043768\\
14.985559393832	15.1857747892736\\
15.1759239845258	14.5458560914609\\
12.2501036660961	11.2402361590881\\
10.7323831250176	12.7950701656599\\
9.36554025073365	9.21371786957834\\
14.7043592283854	14.3063786904862\\
16.1214216821687	15.3338269909757\\
11.6283271412519	10.8481602534177\\
11.9185229626923	13.6992207569787\\
12.0005795457016	9.99152412789111\\
11.488919812771	10.9457334601864\\
8.94489291439293	10.4176800402663\\
12.3619663846535	14.6851875158205\\
11.3342090697278	12.5605690232851\\
20.2871566230906	15.4013624750457\\
11.1343182208339	12.2387418883142\\
13.6909782987699	13.858020230953\\
10.1809971632404	12.0542690392863\\
13.7354213675311	13.8916773719095\\
12.0151824048248	12.9774107680219\\
14.525527917184	12.4105975602477\\
11.0320390133063	11.8216734835595\\
11.0730309419394	9.8025716979592\\
11.9984160294795	12.2508983312195\\
10.3547040945575	11.4208971171799\\
13.7824751835387	17.8596187647283\\
14.2176478685095	14.001186002283\\
9.82381989468399	9.93876356367626\\
12.8586961243628	13.1039952060472\\
12.2833224287795	13.4232469852315\\
13.6788412311972	14.8133700015175\\
13.6597739576154	13.508555642508\\
11.0642524696922	11.9875628275274\\
10.1126690405994	10.2551193638894\\
11.4095045200517	13.3427508737666\\
12.0234610575096	12.3358814590275\\
13.4431434305992	15.2353043517197\\
11.3261310938405	11.710760502201\\
13.4211444741732	14.7819792238427\\
9.7727573973923	10.5964566776407\\
13.2659104270977	15.1521741402932\\
10.7607026115559	11.3352564260239\\
9.55980241672879	10.2380026019067\\
13.5322789200279	14.8120287374728\\
};
\addplot [color=mycolor3,draw opacity=0.5,line width=1.0pt, forget plot]
  table[row sep=crcr]{%
0	0\\
20.2871566230906	20.2871566230906\\
};
\end{axis}
\end{tikzpicture}%

%% file: tex/odgScatter1.tex
%
%
\definecolor{mycolor1}{rgb}{0.06600,0.44300,0.74500}%
\definecolor{mycolor2}{rgb}{0.86600,0.32900,0.00000}%
\definecolor{mycolor3}{rgb}{0.12941,0.12941,0.12941}%
\begin{tikzpicture}

\begin{axis}[%
width=0.45\linewidth,
height=0.4\linewidth,
at={(0in,0in)},
scale only axis,
xmin=-0.978998734887347,
xmax=0.219518257522835,
xlabel style={font=\color{mycolor3}},
xlabel={U-PHAIN-TF ODG},
ymin=-0.759480477032795,
ymax=-3.31716876189603e-10,
ylabel style={font=\color{mycolor3}},
ylabel={Janssen-TF ODG},
axis background/.style={fill=white},
title style={font=\bfseries\color{mycolor3}, at={(0.5in,1in)},},
title={Mask 1},
axis x line*=bottom,
axis y line*=left,
xmajorgrids,
ymajorgrids
]
\addplot[only marks, mark=x, mark options={solid}, mark size=3pt, draw=mycolor1, forget plot] table[row sep=crcr]{%
x	y\\
-1.17539441291115e-06	-0.244813685590607\\
-2.80920069428703e-08	-0.00142562953554304\\
-7.97173798261497e-06	-0.125487768959811\\
-2.97343873967293e-08	-0.000926636575037776\\
-1.41504017037164e-06	-0.0131888043169859\\
-0.000259382201463154	-0.00216900799587449\\
-5.19234362172938e-08	-0.159447958317312\\
-3.82466289750027e-08	-0.00106340389870141\\
-2.44055939901955e-08	-0.0369527091330397\\
-2.10108440867884e-06	-0.0380237218686155\\
-6.7080172527767e-09	-0.00411404278266048\\
-1.15344597872991e-07	-0.00450459873264109\\
-0.000108384289134733	-0.0116524423486517\\
-0.000119985048293358	-0.126883165262669\\
-2.6038514210569e-05	-0.00318495951525222\\
-2.20861544875106e-06	-0.011499490874737\\
-2.67302795009527e-07	-0.00024523102006313\\
-1.31093430155715e-06	-0.000292839888057728\\
-1.99525488646657e-05	-0.0136576745070194\\
-4.52172898235403e-07	-0.011987919324806\\
-6.55620802092471e-08	-0.0111844976993858\\
-3.30340166811993e-05	-0.0842988739659489\\
-6.73862032840589e-09	-0.00279494494779797\\
-8.56322516540331e-05	-0.0112174561732488\\
-4.37739747241039e-05	-0.102766983282024\\
-1.57471821182753e-06	-0.00974274171646172\\
-8.27652435475557e-07	-0.00295691911458107\\
-1.29656836733716e-06	-0.0112954742040081\\
-1.3253186029516e-05	-0.00853976076545493\\
-4.98913060198447e-07	-0.00904700576455042\\
-2.74987190138631e-05	-0.0331349975010475\\
-4.00580709936094e-07	-0.002502009456407\\
-8.62345570951106e-06	-0.00133409348898184\\
-5.23907845106919e-06	-0.0354135609485127\\
-7.93910253094054e-08	-0.0151667786092382\\
-3.31716876189603e-10	-0.000420772631624544\\
-0.000179199931398699	-0.00538118231713369\\
-1.04097785751378e-05	-0.147788805047178\\
-2.3008226435195e-09	-0.00352770816967052\\
-1.54095864957071e-08	-0.0129772675077362\\
-9.18481077860633e-08	-0.0259427744242657\\
-1.57942245948561e-08	-0.00037269805451956\\
-2.84889196677796e-07	-0.0114815883995263\\
-0.000148902223664038	-0.18012499432194\\
-9.62513871058945e-05	-0.0317534494938556\\
-1.79902851016323e-06	-0.0111330170870332\\
-1.94606663939112e-06	-0.0310429285969143\\
-3.96369081556713e-07	-0.00420509018226056\\
-8.72393972173313e-07	-0.014461739171125\\
-3.08579728880432e-07	-0.0654456853245655\\
-2.01812910916033e-06	-0.0524839428148773\\
-1.16134856398276e-07	-0.0130809312468223\\
-1.47319739340901e-06	-0.259480477032795\\
-5.41280670063315e-07	-0.013835632881591\\
-8.70646392741037e-05	-0.016198045610075\\
-1.97616216723873e-08	-0.0681655218970754\\
-3.3743173482037e-06	-0.00521684884311924\\
-0.000242164157963742	-0.00600982104540293\\
-4.25077946175634e-05	-0.0880611870861863\\
-2.6208127970051e-06	-0.0925915867795659\\
};
\addplot [color=mycolor3,draw opacity=0.5, line width=1.0pt, forget plot]
  table[row sep=crcr]{%
-0.759480477032795	-0.759480477032795\\
-3.31716876189603e-10	-3.31716876189603e-10\\
};
\end{axis}
\end{tikzpicture}%

%% file: tex/odgScatter2.tex
%
%
\definecolor{mycolor1}{rgb}{0.06600,0.44300,0.74500}%
\definecolor{mycolor2}{rgb}{0.86600,0.32900,0.00000}%
\definecolor{mycolor3}{rgb}{0.12941,0.12941,0.12941}%
\begin{tikzpicture}

\begin{axis}[%
width=0.45\linewidth,
height=0.4\linewidth,
at={(0in,0in)},
scale only axis,
xmin=-2.56748616593734,
xmax=0.575700446459904,
xlabel style={font=\color{mycolor3}},
xlabel={U-PHAIN-TF ODG},
ymin=-1.99178565135586,
ymax=-6.81215794884338e-08,
ylabel style={font=\color{mycolor3}},
axis background/.style={fill=white},
title style={font=\bfseries\color{mycolor3}, at={(0.5in,1in)},},
title={Mask 2},
axis x line*=bottom,
axis y line*=left,
xmajorgrids,
ymajorgrids
]
\addplot[only marks, mark=x, mark options={solid}, mark size=3pt, draw=mycolor1, forget plot] table[row sep=crcr]{%
x	y\\
-1.53866468934893e-05	-1.49178565135586\\
-8.40280414138306e-08	-0.0617941027027413\\
-6.33336449240574e-05	-0.491312125132662\\
-6.81215794884338e-08	-0.108321077088121\\
-4.74918132553626e-06	-0.0553788179629251\\
-0.00119872717521829	-0.0447035038998607\\
-1.75224819543018e-07	-0.322997899767415\\
-7.45280495095813e-08	-0.034057596974268\\
-1.29849759389344e-07	-0.125452068734329\\
-3.94389812896634e-06	-0.187661863677473\\
-2.7202051455788e-07	-0.0577441806831551\\
-3.54166072469297e-07	-0.104062127663553\\
-0.000413188102676543	-0.417608347273489\\
-0.000817591441549581	-0.446831111056756\\
-1.02846250342736e-05	-0.105172506550922\\
-5.32227242082683e-05	-0.152394149961403\\
-1.93480873633689e-06	-0.0432156499676779\\
-1.55051742396495e-05	-0.0755341344618401\\
-5.45894423531479e-05	-0.0829579008874326\\
-7.2102182002709e-07	-0.0482719686814832\\
-9.27635070979704e-08	-0.126405490159563\\
-0.000139303744909114	-0.461431180230257\\
-1.90707250879996e-07	-0.063186942466448\\
-6.37964168959115e-05	-0.110074372733305\\
-0.000291431597911895	-0.429629914711125\\
-6.07175973144081e-06	-0.181111551230732\\
-7.79920600990636e-07	-0.0440984073864144\\
-8.02183899040187e-06	-0.202729767783005\\
-5.87227323798345e-05	-0.244516557116278\\
-1.62081953192228e-06	-0.266601425516559\\
-0.000286764801249006	-0.140914439467412\\
-6.15399837755604e-07	-0.0404169662137441\\
-7.11073082086955e-05	-0.0325675777405046\\
-1.92087729722346e-05	-0.454891118446511\\
-1.13942554946789e-06	-0.16741515752064\\
-3.38893535456464e-07	-0.0643513055308276\\
-0.00171765431700877	-0.0730964143284858\\
-6.07171627464709e-05	-0.47391420443391\\
-1.47716409770737e-07	-0.156615890074971\\
-1.70906037766372e-07	-0.383278955027844\\
-4.26996621172293e-07	-0.155295604866968\\
-2.36828739730299e-07	-0.0191199618883218\\
-3.27027358082432e-07	-0.115053554193274\\
-0.000343404877277464	-0.332272974402933\\
-0.00100056964583217	-0.521598333290164\\
-1.56264194863809e-05	-0.156363569512024\\
-9.91318060883373e-06	-0.106309021402268\\
-1.07744703470303e-06	-0.0421665101356865\\
-1.61727258074507e-06	-0.0688237352989098\\
-1.1500827064026e-06	-0.377793120914021\\
-2.09483351731876e-05	-0.151228342917467\\
-6.67281430111188e-07	-0.137813379265108\\
-1.5484872282201e-05	-0.718207300418698\\
-3.58568287595062e-06	-0.154312738458135\\
-0.000364881952815921	-0.130186266175553\\
-4.33792926912702e-07	-0.378622437451806\\
-1.02134167114798e-05	-0.0339228083055829\\
-0.00279236528920279	-0.292761255468044\\
-0.000203179021184496	-0.698198334574112\\
-3.18413335556045e-05	-0.242240950122241\\
};
\addplot [color=mycolor3,draw opacity=0.5, line width=1.0pt, forget plot]
  table[row sep=crcr]{%
-1.99178565135586	-1.99178565135586\\
-6.81215794884338e-08	-6.81215794884338e-08\\
};
\end{axis}
\end{tikzpicture}%

%% file: tex/odgScatter3.tex
%
%
\definecolor{mycolor1}{rgb}{0.06600,0.44300,0.74500}%
\definecolor{mycolor2}{rgb}{0.86600,0.32900,0.00000}%
\definecolor{mycolor3}{rgb}{0.12941,0.12941,0.12941}%
\begin{tikzpicture}

\begin{axis}[%
width=0.45\linewidth,
height=0.4\linewidth,
at={(0in,0in)},
scale only axis,
xmin=-3.77552070951219,
xmax=0.814457916190499,
xlabel style={font=\color{mycolor3}},
xlabel={U-PHAIN-TF ODG},
ymin=-2.93482832226646,
ymax=-0.0262344710552256,
ylabel style={font=\color{mycolor3}},
axis background/.style={fill=white},
title style={font=\bfseries\color{mycolor3}, at={(0.5in,1in)},},
title={Mask 3},
axis x line*=bottom,
axis y line*=left,
xmajorgrids,
ymajorgrids
]
\addplot[only marks, mark=x, mark options={solid}, mark size=3pt, draw=mycolor1, forget plot] table[row sep=crcr]{%
x	y\\
-0.0832722830076804	-2.20975627615365\\
-0.0715039534355455	-1.45406871986197\\
-0.0911780563808477	-2.17312805606624\\
-0.0416324414466231	-1.49675052093667\\
-0.0478915491968586	-0.970713489882687\\
-0.0909711302786924	-0.887767689273321\\
-0.0914575851582953	-1.395264069787\\
-0.0791444542397741	-1.23887898433136\\
-0.0729918550356956	-1.0832158253347\\
-0.0707857665737848	-1.87415155103605\\
-0.0582847852499633	-1.32984415541594\\
-0.0297513650583561	-0.811372860886284\\
-0.075437298375352	-1.46534164321717\\
-0.11674584868878	-1.26163574676782\\
-0.0673578635136742	-1.27984138595735\\
-0.0682873995400435	-1.72317675957978\\
-0.0480122370542233	-1.01366990486343\\
-0.0728038822220185	-0.892085554240595\\
-0.158270638911688	-0.639895743137735\\
-0.0388878004801079	-1.41778208966222\\
-0.0295012233049476	-1.13548255826113\\
-0.0814572346666615	-1.63921800408651\\
-0.0741418157976916	-1.4180718011151\\
-0.124816641930821	-2.07976240668023\\
-0.106135579387143	-1.69637277032569\\
-0.0722498066536907	-2.11696995510328\\
-0.100705821378046	-0.874312324374927\\
-0.0496946073501903	-1.07901435405345\\
-0.0804245787215336	-1.52021468924011\\
-0.0637684612067737	-2.07112497120984\\
-0.0941961314091841	-0.997646479112042\\
-0.0749181504951864	-0.60218203664715\\
-0.12484326946473	-0.318022310393868\\
-0.0691378267005156	-1.50699199265459\\
-0.061926229574528	-1.59660237225516\\
-0.0517258226867376	-1.55232005252828\\
-0.0417560725128681	-0.489554642832015\\
-0.0912595578114619	-1.29835053481814\\
-0.0507219872377114	-1.27451175807826\\
-0.0613184194173151	-1.76178744829801\\
-0.090267416811173	-1.65056415522291\\
-0.0377858942228784	-0.29063591834829\\
-0.0637845691466232	-1.90581490050802\\
-0.0469927973389943	-0.871130239282152\\
-0.141707486446602	-2.43482832226646\\
-0.117982224000198	-1.55428179687055\\
-0.0452555007478104	-0.611063410254637\\
-0.0395059794709631	-0.736748279415245\\
-0.0670417345637411	-1.15617239201345\\
-0.0387971955088418	-1.83546451415541\\
-0.0813420220737449	-1.01804833865401\\
-0.0262344710552256	-0.925989118929525\\
-0.064287370263358	-2.14746397051297\\
-0.0739206972419382	-1.15867585399071\\
-0.0518574312122873	-1.08524295656875\\
-0.0624218449332758	-1.80731623633631\\
-0.153318606128106	-0.679613257261359\\
-0.198423089560823	-2.16115996841673\\
-0.0842842050284176	-2.15032146978064\\
-0.109783552424076	-1.33622172904845\\
};
\addplot [color=mycolor3,draw opacity=0.5, line width=1.0pt, forget plot]
  table[row sep=crcr]{%
-2.93482832226646	-2.93482832226646\\
-0.0262344710552256	-0.0262344710552256\\
};
\end{axis}
\end{tikzpicture}%

%% file: tex/odgScatter4.tex
%
%
\definecolor{mycolor1}{rgb}{0.06600,0.44300,0.74500}%
\definecolor{mycolor2}{rgb}{0.86600,0.32900,0.00000}%
\definecolor{mycolor3}{rgb}{0.12941,0.12941,0.12941}%
\begin{tikzpicture}

\begin{axis}[%
width=0.45\linewidth,
height=0.4\linewidth,
at={(0in,0in)},
scale only axis,
xmin=-4.5974085839398,
xmax=0.787100649523499,
xlabel style={font=\color{mycolor3}},
xlabel={U-PHAIN-TF ODG},
ymin=-3.61119137232176,
ymax=-0.199116562094542,
ylabel style={font=\color{mycolor3}},
ylabel={Janssen-TF ODG},
axis background/.style={fill=white},
title style={font=\bfseries\color{mycolor3}, at={(0.5in,1in)},},
title={Mask 4},
axis x line*=bottom,
axis y line*=left,
xmajorgrids,
ymajorgrids
]
\addplot[only marks, mark=x, mark options={solid}, mark size=3pt, draw=mycolor1, forget plot] table[row sep=crcr]{%
x	y\\
-0.941240052496276	-2.64510289914254\\
-1.47842875549705	-2.73309730622114\\
-0.501937518822274	-2.85506742710719\\
-0.932895454869939	-2.67937122426448\\
-1.12637447026268	-2.46921656131414\\
-0.99373325514447	-2.00227149134832\\
-1.34155365643016	-2.67562770738203\\
-1.53307390331859	-2.92547353138177\\
-0.85941906660188	-2.27335561142812\\
-1.59636559758553	-2.61007811192912\\
-1.18928382052316	-2.60902038930297\\
-0.659603637526061	-2.52143230199675\\
-1.25017060523838	-2.72861351798356\\
-0.589062545870991	-1.97218899395568\\
-0.796683779621016	-2.90306280837134\\
-0.542080613511846	-2.58320146215405\\
-1.66496316933286	-2.69655745073667\\
-0.826425591054999	-2.71683570331683\\
-1.56565372606068	-2.79376497961828\\
-0.980864732830796	-2.55068800513764\\
-1.11600505695007	-2.68062816188855\\
-0.710587936769929	-2.76168260826361\\
-1.29094369138553	-2.69459683814014\\
-1.00947557986804	-2.77784627066845\\
-1.20603616962103	-2.58048922798784\\
-1.0740038937656	-3.11119137232176\\
-0.874840809887235	-2.66239145683675\\
-0.893959385187944	-2.20596343557984\\
-0.924477472314422	-2.47847442440082\\
-0.693482580518861	-2.73374360017252\\
-0.706692236345756	-1.94870957659675\\
-0.958453678583581	-2.30344745700343\\
-1.14826377814418	-1.91322667694553\\
-1.13458100566963	-2.41420881707001\\
-0.639412289584008	-2.70959164246681\\
-0.9758885086924	-2.67285448597949\\
-0.199116562094542	-1.30221412518539\\
-1.46658928511244	-2.61826381778971\\
-1.13509824469917	-2.73054643285102\\
-1.38430144415913	-2.66875167803758\\
-1.46405518381546	-2.6411539448759\\
-0.329522472946763	-0.680961731282096\\
-1.12175757558333	-2.57158570954467\\
-1.0195096252186	-2.07223338315983\\
-1.09207442451281	-3.0087052504912\\
-1.30592786451411	-2.22341108346098\\
-0.888482107580039	-2.31717901181797\\
-1.37424177830908	-2.27859484403143\\
-1.25194390373142	-2.76571929226459\\
-0.942002629474874	-2.87842410697979\\
-0.953327226983628	-2.6232172361597\\
-1.35852393352989	-2.54640083023742\\
-1.20907373600452	-2.8849436800425\\
-1.11155301507039	-2.68764941930127\\
-1.04220545093544	-2.37300270294709\\
-0.845063002095507	-2.64722086351729\\
-1.01203499076336	-2.19122007100877\\
-1.24526784412999	-2.64406898895592\\
-0.955250009766729	-2.7342678171431\\
-1.06033169093728	-2.49779988346459\\
};
\addplot [color=mycolor3,draw opacity=0.5, line width=1.0pt, forget plot]
  table[row sep=crcr]{%
-3.61119137232176	-3.61119137232176\\
-0.199116562094542	-0.199116562094542\\
};
\end{axis}
\end{tikzpicture}%

%% file: tex/odgScatter5.tex
%
%
\definecolor{mycolor1}{rgb}{0.06600,0.44300,0.74500}%
\definecolor{mycolor2}{rgb}{0.86600,0.32900,0.00000}%
\definecolor{mycolor3}{rgb}{0.12941,0.12941,0.12941}%
\begin{tikzpicture}

\begin{axis}[%
width=0.45\linewidth,
height=0.4\linewidth,
at={(0in,0in)},
scale only axis,
xmin=-4.62886429079076,
xmax=0.156973961329999,
xlabel style={font=\color{mycolor3}},
xlabel={U-PHAIN-TF ODG},
ymin=-3.75229859991429,
ymax=-0.719591729546472,
ylabel style={font=\color{mycolor3}},
axis background/.style={fill=white},
title style={font=\bfseries\color{mycolor3}, at={(0.5in,1in)},},
title={Mask 5},
axis x line*=bottom,
axis y line*=left,
xmajorgrids,
ymajorgrids
]
\addplot[only marks, mark=x, mark options={solid}, mark size=3pt, draw=mycolor1, forget plot] table[row sep=crcr]{%
x	y\\
-2.02288097680219	-2.99512376435373\\
-2.62728297216304	-3.09779361865251\\
-1.95805186736092	-3.01969179730844\\
-2.31091906922348	-3.01179832215946\\
-2.1855519789272	-2.92772821265433\\
-2.55962877036535	-2.85168808352026\\
-2.54866018514769	-3.07603371812465\\
-2.63213211126444	-3.10985985469734\\
-2.51256780330368	-3.01877079659071\\
-2.32969703068454	-2.94268003390099\\
-2.3936931755397	-2.95197161035179\\
-2.15875856912077	-2.79742009210305\\
-1.69666910919636	-3.03583036256426\\
-2.44646334493457	-2.97876732375537\\
-2.68710265939532	-3.25229859991429\\
-1.93533800450424	-2.99007179723286\\
-2.11922660260531	-3.02671822265726\\
-2.40677002447412	-3.11196148133256\\
-2.75737922399367	-3.02182772856718\\
-2.49251166020749	-2.94156580077012\\
-2.22599893455349	-2.88610296924811\\
-1.02912225696229	-2.37390060632174\\
-2.32481155507652	-3.02229026136903\\
-2.41824028572389	-3.09234677108388\\
-1.7128403660212	-2.90923444472035\\
-2.29836293973109	-3.19026168181555\\
-2.44857596679191	-3.00257230740563\\
-2.47040320402645	-3.01115276809216\\
-2.49775101181003	-2.92689609361382\\
-2.09682854114245	-3.01815441611417\\
-1.79739622966106	-2.67454266794056\\
-2.53925599404395	-3.04778061440792\\
-2.30887480574134	-2.22728749563077\\
-2.61673914112904	-3.12427460455882\\
-2.43899682194171	-2.9932357986806\\
-2.44973050721568	-3.10161385037747\\
-0.719591729546472	-2.44457936894621\\
-2.37937717294946	-2.97083834718206\\
-2.5679581598704	-3.05617062094907\\
-2.25890185718698	-3.03438127719128\\
-2.51375036914163	-3.11818922016886\\
-1.17164374981965	-2.25704823118022\\
-2.56391435977359	-3.00025384431432\\
-2.41219524457297	-2.90081233020871\\
-1.98426370655983	-3.05943703695735\\
-2.2947853663364	-2.98476313614844\\
-2.37344889089931	-2.81018433687929\\
-2.16008491414095	-2.96328618614459\\
-2.62015358584534	-2.98847274133902\\
-2.43044658204393	-3.07707493815497\\
-2.23829628513315	-2.9626608052873\\
-2.78516547804528	-3.03173060497229\\
-2.16244947570148	-3.13677846054426\\
-2.63107980370171	-2.97944432023508\\
-2.23223799031174	-2.83826044653119\\
-2.35137974839005	-3.09426714763017\\
-2.32257927195565	-2.87032884400216\\
-2.39018561984121	-2.95214277878512\\
-2.60358752638022	-3.03420064292538\\
-2.3015424564008	-2.91437191609327\\
};
\addplot [color=mycolor3,draw opacity=0.5, line width=1.0pt, forget plot]
  table[row sep=crcr]{%
-3.75229859991429	-3.75229859991429\\
-0.719591729546472	-0.719591729546472\\
};
\end{axis}
\end{tikzpicture}%

%% file: tex/odgScatter6.tex
%
%
\definecolor{mycolor1}{rgb}{0.06600,0.44300,0.74500}%
\definecolor{mycolor2}{rgb}{0.86600,0.32900,0.00000}%
\definecolor{mycolor3}{rgb}{0.12941,0.12941,0.12941}%
\begin{tikzpicture}

\begin{axis}[%
width=0.45\linewidth,
height=0.4\linewidth,
at={(0in,0in)},
scale only axis,
xmin=-4.39908845307144,
xmax=-1.09622784397708,
xlabel style={font=\color{mycolor3}},
xlabel={U-PHAIN-TF ODG},
ymin=-3.79414233095866,
ymax=-1.70117396608986,
ylabel style={font=\color{mycolor3}},
axis background/.style={fill=white},
title style={font=\bfseries\color{mycolor3}, at={(0.5in,1in)},},
title={Mask 6},
axis x line*=bottom,
axis y line*=left,
xmajorgrids,
ymajorgrids
]
\addplot[only marks, mark=x, mark options={solid}, mark size=3pt, draw=mycolor1, forget plot] table[row sep=crcr]{%
x	y\\
-2.52737428244902	-3.05380385950775\\
-2.7172642175552	-3.16429571895094\\
-2.30177024303451	-3.02040142412712\\
-2.78721003398635	-3.19476684136964\\
-2.86555103043875	-3.06013164780177\\
-2.88201474375386	-3.1201399262488\\
-2.88780939063404	-3.23611612916762\\
-2.59193087046995	-3.1747380332745\\
-2.79428283585701	-3.12722008043805\\
-2.65139253257167	-3.10493089780769\\
-2.47469099579839	-3.20283225247066\\
-2.60974953219049	-3.04853070893834\\
-2.48045683659075	-3.12166151950061\\
-2.51140340489353	-3.01605206618548\\
-2.23112956148606	-3.21049165505137\\
-2.52072218584113	-3.17801119571003\\
-2.6169696503357	-3.25761326648559\\
-2.89420206147005	-3.13818881328836\\
-2.97438923305513	-3.1803321475615\\
-2.87600461078422	-3.06917829758413\\
-2.61148902964467	-3.01363308162475\\
-1.99502389122571	-2.95439255627825\\
-2.6637949659111	-3.20438329160607\\
-2.59469402801531	-3.18945474698421\\
-2.54395531582791	-3.00322285209391\\
-2.77906902279915	-3.21420751701559\\
-2.62453124384964	-3.07872658442151\\
-2.92900832000379	-3.17455645150262\\
-2.80802305747342	-3.12590116913692\\
-2.50389599043212	-3.15974219223432\\
-2.33718859195125	-2.95480518704572\\
-2.90447364500036	-3.15014771546149\\
-2.75376820226429	-3.01857669324642\\
-2.84986251158025	-3.11883962199583\\
-2.82293807886308	-3.29414233095866\\
-2.92786107426426	-3.20466483247339\\
-1.70117396608986	-2.84600263547398\\
-2.69901207892114	-3.02092697170313\\
-2.77125186588304	-3.16198127205053\\
-2.84922884784693	-3.20631305327204\\
-2.75385403227203	-3.21992672718292\\
-2.25945138316763	-2.46596745564134\\
-2.72371915024264	-3.08483338415557\\
-2.36746237538724	-2.95439402180796\\
-2.39882082211428	-3.12974805874589\\
-2.61862733591386	-3.07094360662887\\
-2.88716371801203	-3.02517670601437\\
-2.68122312842051	-3.14138513238303\\
-2.92218609821806	-3.12119551631366\\
-2.58970289273521	-3.16638791821385\\
-2.64009054951615	-3.10580783198022\\
-2.8515350694852	-3.1973388019258\\
-2.55435128806429	-3.17250637746358\\
-2.83466107962663	-3.19986078870769\\
-2.55172296777507	-3.04679343099311\\
-2.73382150496727	-3.11682972253117\\
-2.63263359164145	-3.10616557853695\\
-2.70686596047192	-3.03505000608735\\
-2.82604087310252	-3.1467650633664\\
-2.49344009380315	-3.06168632136756\\
};
\addplot [color=mycolor3,draw opacity=0.5, line width=1.0pt, forget plot]
  table[row sep=crcr]{%
-3.79414233095866	-3.79414233095866\\
-1.70117396608986	-1.70117396608986\\
};
\end{axis}
\end{tikzpicture}%

%% file: tex/listeningNew.tex
%
%

\definecolor{mycolor1}{rgb}{0.76650,0.86075,0.93625}%
\definecolor{mycolor2}{rgb}{0.06600,0.44300,0.74500}%
\definecolor{mycolor3}{rgb}{0.86600,0.32900,0.00000}%
\definecolor{mycolor4}{rgb}{0.12941,0.12941,0.12941}%

\definecolor{phain}{rgb}{0.00000,0.44700,0.74100}%
\definecolor{dpai}{rgb}
{0.92900,0.69400,0.12500}%
\definecolor{janssen}{rgb}{0.49400,0.18400,0.55600}%
\definecolor{reference}{rgb}{0.4660,0.6740,0.1880}%
\definecolor{anchor}{rgb}{0.6350,0.0780,0.1840}%

\begin{tikzpicture}

\begin{axis}[%
width=1\linewidth,
height=0.7\linewidth,
at={(0in,0in)},
scale only axis,
unbounded coords=jump,
xmin=0.5,
xmax=5.5,
xtick={1,2,3,4,5},
xticklabels={{anchor},{reference},{Janssen-TF},{DPAI},{U-PHAIN-TF}},
ymin=-5,
ymax=105,
ylabel style={font=\color{mycolor4}},
ylabel={MUSHRA score},
axis background/.style={fill=white},
ymajorgrids,
yminorgrids,
major grid style={white!75!black},
minor grid style={white!90!black},
minor y tick num=3,
]

\addplot[area legend, draw=black,draw opacity=0.1, fill=anchor, fill opacity=0.3, forget plot]
table[row sep=crcr] {%
x	y\\
0.875	0\\
0.75	0.0898979384872773\\
1.25	0.0898979384872773\\
1.125	0\\
1.25	0\\
0.75	0\\
0.875	0\\
}--cycle;

\addplot[area legend, draw=black,draw opacity=0.1, fill=phain, fill opacity=0.3, forget plot]
table[row sep=crcr] {%
x	y\\
4.875	96\\
4.75	98.359820885291\\
5.25	98.359820885291\\
5.125	96\\
5.25	93.640179114709\\
4.75	93.640179114709\\
4.875	96\\
}--cycle;

\addplot[area legend, draw=black,draw opacity=0.1, fill=janssen, fill opacity=0.3, forget plot]
table[row sep=crcr] {%
x	y\\
2.875	82\\
2.75	85.8656113549529\\
3.25	85.8656113549529\\
3.125	82\\
3.25	78.1343886450471\\
2.75	78.1343886450471\\
2.875	82\\
}--cycle;

\addplot[area legend, draw=black,draw opacity=0.1, fill=dpai, fill opacity=0.3, forget plot]
table[row sep=crcr] {%
x	y\\
3.875	66\\
3.75	69.0790043931893\\
4.25	69.0790043931893\\
4.125	66\\
4.25	62.9209956068107\\
3.75	62.9209956068107\\
3.875	66\\
}--cycle;
\addplot [color=anchor, dashed, forget plot]
  table[row sep=crcr]{%
1	1\\
1	2\\
};
\addplot [color=reference, dashed, forget plot]
  table[row sep=crcr]{%
2	100\\
2	100\\
};
\addplot [color=janssen, dashed, forget plot]
  table[row sep=crcr]{%
3	100\\
3	100\\
};
\addplot [color=dpai, dashed, forget plot]
  table[row sep=crcr]{%
4	83.25\\
4	100\\
};
\addplot [color=phain, dashed, forget plot]
  table[row sep=crcr]{%
5	100\\
5	100\\
};
\addplot [color=anchor, dashed, forget plot]
  table[row sep=crcr]{%
1	0\\
1	0\\
};
\addplot [color=reference, dashed, forget plot]
  table[row sep=crcr]{%
2	100\\
2	100\\
};
\addplot [color=janssen, dashed, forget plot]
  table[row sep=crcr]{%
3	0\\
3	57\\
};
\addplot [color=dpai, dashed, forget plot]
  table[row sep=crcr]{%
4	0\\
4	49\\
};
\addplot [color=phain, dashed, forget plot]
  table[row sep=crcr]{%
5	39\\
5	73.75\\
};
\addplot [color=anchor, forget plot]
  table[row sep=crcr]{%
0.875	2\\
1.125	2\\
};
\addplot [color=reference, forget plot]
  table[row sep=crcr]{%
1.875	100\\
2.125	100\\
};
\addplot [color=janssen, forget plot]
  table[row sep=crcr]{%
2.875	100\\
3.125	100\\
};
\addplot [color=dpai, forget plot]
  table[row sep=crcr]{%
3.875	100\\
4.125	100\\
};
\addplot [color=phain, forget plot]
  table[row sep=crcr]{%
4.875	100\\
5.125	100\\
};
\addplot [color=anchor, forget plot]
  table[row sep=crcr]{%
0.875	0\\
1.125	0\\
};
\addplot [color=reference, forget plot]
  table[row sep=crcr]{%
1.875	100\\
2.125	100\\
};
\addplot [color=janssen, forget plot]
  table[row sep=crcr]{%
2.875	0\\
3.125	0\\
};
\addplot [color=dpai, forget plot]
  table[row sep=crcr]{%
3.875	0\\
4.125	0\\
};
\addplot [color=phain, forget plot]
  table[row sep=crcr]{%
4.875	39\\
5.125	39\\
};
\addplot [color=anchor, forget plot]
  table[row sep=crcr]{%
0.875	0\\
0.75	0.0898979384872773\\
0.75	1\\
1.25	1\\
1.25	0.0898979384872773\\
1.125	0\\
1.25	0\\
1.25	0\\
0.75	0\\
0.75	0\\
0.875	0\\
};
\addplot [color=reference, forget plot]
  table[row sep=crcr]{%
1.875	100\\
1.75	100\\
1.75	100\\
2.25	100\\
2.25	100\\
2.125	100\\
2.25	100\\
2.25	100\\
1.75	100\\
1.75	100\\
1.875	100\\
};
\addplot [color=janssen, forget plot]
  table[row sep=crcr]{%
2.875	82\\
2.75	85.8656113549529\\
2.75	100\\
3.25	100\\
3.25	85.8656113549529\\
3.125	82\\
3.25	78.1343886450471\\
3.25	57\\
2.75	57\\
2.75	78.1343886450471\\
2.875	82\\
};
\addplot [color=dpai, forget plot]
  table[row sep=crcr]{%
3.875	66\\
3.75	69.0790043931893\\
3.75	83.25\\
4.25	83.25\\
4.25	69.0790043931893\\
4.125	66\\
4.25	62.9209956068107\\
4.25	49\\
3.75	49\\
3.75	62.9209956068107\\
3.875	66\\
};
\addplot [color=phain, forget plot]
  table[row sep=crcr]{%
4.875	96\\
4.75	98.359820885291\\
4.75	100\\
5.25	100\\
5.25	98.359820885291\\
5.125	96\\
5.25	93.640179114709\\
5.25	73.75\\
4.75	73.75\\
4.75	93.640179114709\\
4.875	96\\
};
\addplot [color=anchor, line width=1.0pt, forget plot]
  table[row sep=crcr]{%
0.875	0\\
1.125	0\\
};
\addplot [color=reference, line width=1.0pt, forget plot]
  table[row sep=crcr]{%
1.875	100\\
2.125	100\\
};
\addplot [color=janssen, line width=1.0pt, forget plot]
  table[row sep=crcr]{%
2.875	82\\
3.125	82\\
};
\addplot [color=dpai, line width=1.0pt, forget plot]
  table[row sep=crcr]{%
3.875	66\\
4.125	66\\
};
\addplot [color=phain, line width=1.0pt, forget plot]
  table[row sep=crcr]{%
4.875	96\\
5.125	96\\
};
\addplot [color=anchor, only marks, mark=+,mark size=3pt, mark options={solid, draw=anchor}, forget plot]
  table[row sep=crcr]{%
1	3\\
1	4\\
1	4\\
1	4\\
1	6\\
1	9\\
1	9\\
1	10\\
1	11\\
1	11\\
1	14\\
1	14\\
1	14\\
1	15\\
1	15\\
1	15\\
1	15\\
1	15\\
1	15\\
1	15\\
1	17\\
1	18\\
1	18\\
1	19\\
1	19\\
1	19\\
1	19\\
1	20\\
1	20\\
1	20\\
1	20\\
1	20\\
1	20\\
1	20\\
1	20\\
1	20\\
1	20\\
1	20\\
1	20\\
1	21\\
1	21\\
1	21\\
1	21\\
1	22\\
1	22\\
1	22\\
1	23\\
1	23\\
1	23\\
1	24\\
1	24\\
1	24\\
1	24\\
1	25\\
1	25\\
1	25\\
1	25\\
1	25\\
1	25\\
1	25\\
1	25\\
1	25\\
1	25\\
1	25\\
1	25\\
1	25\\
1	25\\
1	25\\
1	25\\
1	27\\
1	27\\
1	31\\
1	36\\
1	36\\
};
\addplot [color=reference, only marks, mark=+,mark size=3pt, mark options={solid, draw=reference}, forget plot]
  table[row sep=crcr]{%
2	64\\
2	71\\
2	79\\
2	80\\
2	81\\
2	82\\
2	84\\
2	88\\
2	89\\
2	90\\
2	90\\
2	90\\
2	90\\
2	90\\
2	91\\
2	92\\
2	94\\
2	94\\
2	95\\
2	95\\
2	96\\
2	97\\
2	98\\
};
\addplot [color=phain, only marks, mark=+,mark size=3pt, mark options={solid,  draw=phain}, forget plot]
  table[row sep=crcr]{%
5	0\\
5	0\\
5	0\\
5	0\\
5	0\\
5	22\\
5	22\\
5	25\\
5	25\\
5	27\\
5	29\\
5	30\\
5	31\\
5	33\\
5	33\\
};
\end{axis}

\end{tikzpicture}%

%% file: tex/irmas_snr_alt.tex
%
%
\definecolor{mycolor1}{rgb}{0.00000,0.44700,0.74100}%
\definecolor{mycolor2}{rgb}{0.85000,0.32500,0.09800}%
\definecolor{mycolor3}{rgb}{0.00000,0.22350,0.5}
\definecolor{mycolor4}{rgb}{0.49400,0.18400,0.55600}%
\begin{tikzpicture}

\begin{axis}[%
width=0.45\linewidth,
height=0.45\linewidth,
at={(0in,0in)},
scale only axis,
xmin=1,
xmax=6,
xtick={1, 2, 3, 4, 5, 6},
xlabel style={font=\color{white!15!black}},
xlabel={gap length},
ymin=0,
ymax=100,
ylabel style={font=\color{white!15!black}},
ylabel={$\SNRfull$ [dB]},
axis background/.style={fill=white},
xmajorgrids,
ymajorgrids,
legend style={legend cell align=left, align=left, draw=white!15!black, font=\footnotesize}
]
\addplot[area legend, draw=black, fill=mycolor1, draw opacity=0.1, fill opacity=0.2, forget plot]
table[row sep=crcr] {%
x	y\\
1	92.2214825141404\\
2	88.8916081992955\\
3	32.0329423996636\\
4	19.5383781820734\\
5	14.2984708925162\\
6	11.847310733173\\
6	12.7476223051282\\
5	15.6751574249208\\
4	21.2099969215164\\
3	33.8860212997576\\
2	89.7051044409751\\
1	93.6637056387268\\
}--cycle;
\addplot [color=mycolor1, line width=1.0pt]
  table[row sep=crcr]{%
1	92.9425940764336\\
2	89.2983563201353\\
3	32.9594818497106\\
4	20.3741875517949\\
5	14.9868141587185\\
6	12.2974665191506\\
};
\addlegendentry{Proposed}



\addplot[area legend, draw=black, fill=mycolor3, draw opacity=0.1, fill opacity=0.2, forget plot]
table[row sep=crcr] {%
x	y\\
1	28.1841676987445\\
2	19.4137263462179\\
3	13.911079599001\\
4	11.30570009299\\
5	8.68936100317983\\
6	7.67881032748048\\
6	8.17457478647402\\
5	9.27621085866414\\
4	11.8972909990999\\
3	14.8263650248332\\
2	20.5980810361501\\
1	29.3843056696789\\
}--cycle;
\addplot [color=mycolor3, dashed, line width=1.0pt]
  table[row sep=crcr]{%
1	28.7842366842117\\
2	20.005903691184\\
3	14.3687223119171\\
4	11.6014955460449\\
5	8.98278593092199\\
6	7.92669255697725\\
};
\addlegendentry{Alternative}

\end{axis}

\end{tikzpicture}%

%% file: tex/irmas_odg_alt.tex
%
%
\definecolor{mycolor1}{rgb}{0.00000,0.44700,0.74100}%
\definecolor{mycolor2}{rgb}{0.85000,0.32500,0.09800}%
\definecolor{mycolor3}{rgb}{0.00000,0.22350,0.5}
\definecolor{mycolor4}{rgb}{0.49400,0.18400,0.55600}%
\begin{tikzpicture}

\begin{axis}[%
width=0.45\linewidth,
height=0.45\linewidth,
at={(0in,0in)},
scale only axis,
xmin=1,
xmax=6,
xtick={1, 2, 3, 4, 5, 6},
xlabel style={font=\color{white!15!black}},
xlabel={gap length},
ymin=-3.5,
ymax=0,
ylabel style={font=\color{white!15!black}},
ylabel={ODG},
axis background/.style={fill=white},
xmajorgrids,
ymajorgrids,
]

\addplot[area legend, draw=black, fill=mycolor1, draw opacity=0.1, fill opacity=0.2, forget plot]
table[row sep=crcr] {%
x	y\\
1	-3.89969348906207e-05\\
2	-0.000270485193450364\\
3	-0.0834994580693135\\
4	-1.10756023736087\\
5	-2.36552958948899\\
6	-2.69926358844513\\
6	-2.59720094142088\\
5	-2.19861151202215\\
4	-0.975578830234344\\
3	-0.0689792592551805\\
2	-6.95209261367616e-05\\
1	-1.41744219030295e-05\\
}--cycle;
\addplot [color=mycolor1, line width=1.0pt]
  table[row sep=crcr]{%
1	-2.65856783968251e-05\\
2	-0.000170003059793563\\
3	-0.076239358662247\\
4	-1.04156953379761\\
5	-2.28207055075557\\
6	-2.648232264933\\
};



\addplot[area legend, draw=black, fill=mycolor3, draw opacity=0.1, fill opacity=0.2, forget plot]
table[row sep=crcr] {%
x	y\\
1	-0.17795507836556\\
2	-1.22932611410282\\
3	-2.38146511930874\\
4	-2.81097895579828\\
5	-3.02504332087371\\
6	-3.09096499205286\\
6	-3.04757567117441\\
5	-2.97018935552167\\
4	-2.72381996763708\\
3	-2.24669855338817\\
2	-1.10090456424314\\
1	-0.156510407882705\\
}--cycle;
\addplot [color=mycolor3, dashed, line width=1.0pt]
  table[row sep=crcr]{%
1	-0.167232743124133\\
2	-1.16511533917298\\
3	-2.31408183634846\\
4	-2.76739946171768\\
5	-2.99761633819769\\
6	-3.06927033161363\\
};

\end{axis}

\end{tikzpicture}%

%% file: tex/thresholding.tex
%
%
\definecolor{mycolor1}{rgb}{0.06600,0.44300,0.74500}%
\definecolor{mycolor2}{rgb}{0.86600,0.32900,0.00000}%
\definecolor{mycolor3}{rgb}{0.92900,0.69400,0.12500}%
\definecolor{mycolor4}{rgb}{0.52100,0.08600,0.81900}%
\definecolor{mycolor5}{rgb}{0.23100,0.66600,0.19600}%
\definecolor{mycolor6}{rgb}{0.18400,0.74500,0.93700}%
\definecolor{mycolor7}{rgb}{0.81900,0.01500,0.54500}%
\definecolor{mycolor8}{rgb}{0.12941,0.12941,0.12941}%
\begin{tikzpicture}

\begin{axis}[%
width=1\linewidth,
height=0.7\linewidth,
at={(0in,0in)},
scale only axis,
xmin=1,
xmax=6,
xtick={1, 2, 3, 4, 5, 6},
xlabel style={font=\color{mycolor8}},
xlabel={gap length},
ymin=-3,
ymax=0,
ylabel style={font=\color{mycolor8}},
ylabel={ODG},
axis background/.style={fill=white},
xmajorgrids,
ymajorgrids,
legend style={at={(0.03,0.03)}, anchor=south west, legend cell align=left, align=left, draw=white!15!black,font=\footnotesize}
]
\addplot [color=mycolor1, mark=o,mark size=2pt,line width=1.0pt, mark options={solid, mycolor1}, draw opacity=0.5]
  table[row sep=crcr]{%
1	-0.00137411106744478\\
2	-0.00143463580323733\\
3	-0.0974601484526003\\
4	-1.1533786895909\\
5	-2.5377863011959\\
6	-2.91871447054283\\
};
\addlegendentry{B-PHAIN-TF}

\addplot [color=mycolor1, mark=x,mark size=3pt,line width=1.0pt, mark options={solid, mycolor1}]
  table[row sep=crcr]{%
1	-0.00137406088474368\\
2	-0.00139682261660168\\
3	-0.0786438645937406\\
4	-1.04649984600058\\
5	-2.22672318120024\\
6	-2.43479406385542\\
};
\addlegendentry{U-PHAIN-TF}

\addplot [color=mycolor6,dashed,line width=1.0pt ]
  table[row sep=crcr]{%
1	-0.00137412948987237\\
2	-0.0013891825417951\\
3	-0.0342706703517024\\
4	-0.574962011680164\\
5	-1.55411331516501\\
6	-2.0747274866991\\
};
\addlegendentry{B-PHAIN-TF (oracle)}

\addplot [color=mycolor4, mark=asterisk,mark size=3pt,line width=1.0pt, mark options={solid, mycolor4},draw opacity = 0.75]
  table[row sep=crcr]{%
1	-0.00137423131517656\\
2	-0.00137423131517656\\
3	-0.0898895520132541\\
4	-1.32446522517955\\
5	-2.35778075237737\\
6	-2.54831426899658\\
};
\addlegendentry{U-PHAIN-TF$_{\ell_2}$}

\addplot [color=mycolor5, mark=x,mark size=3pt,line width=1.0pt, mark options={solid, mycolor5}, draw opacity = 0.75]
  table[row sep=crcr]{%
1	-0.00137423131517656\\
2	-0.00137427761703623\\
3	-0.0776140132784056\\
4	-1.28486806616362\\
5	-2.39107181629091\\
6	-2.52235514632206\\
};
\addlegendentry{U-PHAIN-TF$_{\ell_2^2}$}

\addplot [color=mycolor3, mark=diamond, mark size=2pt,line width=1.0pt, mark options={solid, mycolor3},draw opacity=0.75]
  table[row sep=crcr]{%
1	-0.0012835153198596\\
2	-0.00310872309117505\\
3	-0.209877993887388\\
4	-1.3374027124558\\
5	-2.2695784768696\\
6	-2.51119964105297\\
};
\addlegendentry{U-PHAIN-TF$_{\mathrm{SH}}$}

\addplot [color=mycolor7,dashed ,mark=+, mark size=3pt,line width=1.0pt, mark options={solid, mycolor7}, draw opacity=0.75]
  table[row sep=crcr]{%
1	-0.00137406088474368\\
2	-0.00139686287313978\\
3	-0.0735647456794437\\
4	-1.01634347379358\\
5	-2.23862873631603\\
6	-2.53483721146067\\
};
\addlegendentry{U-PHAIN-TF$_{\mathrm{shrink}_p}$}

\end{axis}

\end{tikzpicture}%

%% file: tex/normalization.tex
%
%
\definecolor{mycolor1}{rgb}{0.00000,0.44700,0.74100}%
\definecolor{mycolor2}{rgb}{0.85000,0.32500,0.09800}%
\definecolor{mycolor3}{rgb}{0.49400,0.18400,0.55600}%
\begin{tikzpicture}

\begin{axis}[%
width=1\linewidth,
height=0.7\linewidth,
at={(0in,0in)},
scale only axis,
xmode=log,
xmin=1e-07,
xmax=100,
xminorticks=true,
xlabel style={font=\color{white!15!black}},
xlabel={lambdas},
ymin=-1.7,
ymax=-0.9,
ylabel style={font=\color{white!15!black}},
ylabel={ODG},
axis background/.style={fill=white},
xmajorgrids,
ymajorgrids,
legend style={at={(0.03,0.03)}, anchor=south west, legend cell align=left, align=left, draw=white!15!black}
]
\addplot [color=mycolor1,dashed,draw opacity=0.75, line width=1.0pt, mark=x,mark size=4pt, mark options={solid, mycolor1}]
  table[row sep=crcr]{%
1e-07	-1.60573706204954\\
1e-06	-1.6030068798259\\
1e-05	-1.5457796952626\\
0.0001	-1.39209050126374\\
0.001	-1.01208282104633\\
0.01	-0.991431841915339\\
0.1	-1.12221877798049\\
1	-1.28367956741604\\
10	-1.49892873951576\\
100	-1.6212552275481\\
};
\addlegendentry{Test with original amplitude}


\addplot [color=mycolor1, line width=1.0pt, mark=x,mark size=4pt, mark options={solid, mycolor1}]
  table[row sep=crcr]{%
1e-07	-1.60601231309408\\
1e-06	-1.60561513852348\\
1e-05	-1.58540090191659\\
0.0001	-1.48530047188308\\
0.001	-1.24748910546481\\
0.01	-0.964798025250955\\
0.1	-1.04336662788788\\
1	-1.17246603962212\\
10	-1.39219267102956\\
100	-1.60704827670719\\
};
\addlegendentry{Test with normalized amplitude}

\end{axis}

\end{tikzpicture}%